\documentclass[preprint]{aastex63}
\usepackage{graphicx,color}
\usepackage{breakurl, hyperref}
\usepackage{hyperref}
\def\UrlBreaks{\do\/\do\-}
\shortauthors{Allers \& Liu}
\shorttitle{$W$-band Filter: Design and Pilot Survey}

\newcommand\note[1]{\textbf{\color{red}#1}}   

\begin{document}

\newcommand{\wfilt}{\mbox{$W$}}
\newcommand{\degs}{\mbox{$^{\circ}$}}
\newcommand{\mjup}{\mbox{$M_{\rm{Jup}}$}}
\newcommand{\etal}{et al.}
\newcommand{\eg}{e.g.}
\newcommand{\ie}{i.e.}
\newcommand{\htwoo}{{\hbox{H$_2$O}}} 
\newcommand{\units}[1]{\,\mbox{#1}}  

\newcommand{\HST}{{\sl HST}}
\newcommand{\TBD}{\note{TBD}}

\title{A Novel Survey for Young Substellar Objects with the $W$-band Filter.\\
  I. Filter Design and New Discoveries in Ophiuchus and Perseus}
\author[0000-0003-0580-7244]{K.~N.~Allers}
\affil{Department of Physics and Astronomy, Bucknell University, Lewisburg, PA 17837, USA}
\affil{Visiting Astronomer at the Infrared Telescope Facility, which is operated by the University of Hawaii under contract NNH14CK55B with the National Aeronautics and Space Administration.} 
\author[0000-0003-2232-7664]{Michael.~C.~Liu}
\affil{Institute for Astronomy, University of Hawai'i,
2680 Woodlawn Drive, Honolulu, HI 96822, USA}

  
\begin{abstract}

  We present the design and implementation of a medium-band near-IR
  filter tailored for detecting low-mass stars and brown dwarfs from the
  summit of Maunakea. The $W$-band filter is centered at 1.45~\micron\
  with a bandpass width of 6\%, designed to measure the depth of the
  \htwoo\ water absorption prominent in objects with spectral types of
  M6 and later. When combined with standard $J$ and $H$ photometry, the
  $W$-band filter is designed to determine spectral types to
  $\approx$1.4~subtypes for late-M and L~dwarfs, largely
  independent of surface gravity and reddening. This filter's primary
  application is completing the census of young substellar objects in
  star-forming regions, using $W$-band selection to greatly reduce
  contamination by reddened background stars that impede broad-band
  imaging surveys. We deployed the filter on the UH 88 inch telescope to
  survey $\sim$3~degree$^2$ of the NGC 1333, IC~348, and $\rho$
  Ophiuchus star-forming regions. Our spectroscopic followup of $W$-band
  selected candidates resulted in the confirmation of 48 ultracool
  dwarfs with a success rate of 89\%, demonstrating the efficacy of this
  new filter and selection method.

\end{abstract}

\section{Introduction} \label{intro}

Broad-band filters in ground-based optical/infrared astronomy are
designed to maximize sensitivity, with bandpasses placed in the regions
of highest atmospheric transmission and spanning a large continuous
swath of wavelengths \citep[e.g.][]{2005ARA&A..43..293B}. As a result,
broad-band photometric systems have a small total number of filters
($\approx$5), meaning multi-band images provide some basic spectral
information on astronomical sources. It has long been appreciated that
in addition to considering the Earth's atmosphere, consideration of the
intrinsic spectrum of astrophysical sources during the design process
can result in filters tailored for specific applications.

Such tailored filters typically sacrifice sensitivity and broad
wavelength coverage in exchange for greater information content in the
resulting photometry. Such filters are often used to detect rare classes
of objects on the sky, where broad-band photometry does not sufficiently
distinguish objects of interest from a much larger sea of undesired
stars and galaxies. Well-known examples include the Str{\"o}mgren
filters for measuring temperature, surface gravity, and reddening of
early-type stars \citep[][]{1966ARA&A...4..433S} and the methane-band
filters for detection and spectral typing of T~dwarfs
\citep{1996Natur.384..243R, 2005AJ....130.2326T}. The young L~dwarf
  companion G~196-3B was found by \citet{1998Sci...282.1309R} using
  narrow-band optical filters described by \citet{1996ApJ...469..706M}.

\citet{najita00} pioneered the application of tailored filters for young
low-mass stars and brown dwarfs. With spectral types of late-M and later
(and thus commonly referred to as ultracool objects), these objects have
ample water absorption in their photospheres, which is manifested in
their spectra as deep, broad absorption troughs, analogous to the
telluric bands that demarcate the traditional $JHK$ broad-band filters.
Najita \etal\ used a combination of 3 narrow-band
($\Delta\lambda/\lambda\approx1\%$) filters in the Hubble Space Telescope (HST)/ NICMOS imager
(F166N, F190N, and F215N) to measure the 1.9~\micron\ water
absorption for objects in the IC~348 star-forming region. The first and
third filters are analogous to the $H$ and $K$~filters and measure the
bright continuum flux, while the F190N filter resides in the water band.
To derive spectral types for all sources in their images, they formed a
reddening-independent $Q$ index \citep{1953ApJ...117..313J}, calibrated
from IC~348 sources with spectral types previously measured
spectroscopically. This resulted in a spectral-typing precision of
$\approx$1~subtype for objects with water absorption. (This method
cannot classify earlier-type sources, which lack water absorption, but
can distinguish such objects from later-type ones.) They estimate
reddenings for objects in their HST photometry from the two resulting
colors (F166N-F190N and F190N-F215N). Their work resulted in a deep
measurement of the initial mass function (IMF) in this cluster, as well
as enabling the first robust survey for circumstellar disks at such low
masses \citep{2003ApJ...585..372L}. A somewhat similar approach for
ground-based and aircraft-based observations based on medium-band IR
filters was developed by \citealp{2003ApJ...597..555M}.

The shape and low-mass cutoff of the IMF are signatures of the star (and
brown dwarf) formation process. Measuring these properties and their
dependence on environment provides the empirical foundation for testing
theoretical models. In recent years, wide-field optical and infrared
imagers have pushed photometric surveys of star-forming regions ever
deeper \citep[e.g.][]{alves10, scholz12} and thus to lower masses, down
to $\approx$5~\mjup\ \citep[e.g.][]{2013MNRAS.435.2474L,
  2017ApJ...842...65Z}. Using broad-band filters, such surveys readily
identify hundreds of brown-dwarf candidates in young ($\sim$Myr)
star-forming regions. The largest obstacle to this approach is its need
for spectroscopic followup, since deeper imaging also means an increase
in contaminating sources (mostly reddened background stars). Because
spectroscopy of large samples is telescope-time intensive, strict
photometric selection criteria are commonly used to reduce the number of
contaminants in their samples. Such criteria may be problematic, by
relying on evolutionary model isochrones that have not been validated at
such young ages and low masses. Similarly, recent studies of
$\sim$10--100~Myr old field brown dwarfs find that objects can be
fainter in the near-IR than expected \citep[e.g.][]{2013AN....334...85L,
  2016ApJS..225...10F, 2016ApJ...833...96L}.
So altogether,
broad-band surveys may have undesirable and/or uncertain incompleteness
in the low-mass regime. Relaxing the photometric selection criteria to
increase completeness is not a practical solution, since the increased
number of candidates further strains follow-up spectroscopy \citep{barsony12}.

As demonstrated by Najita \etal\ (2000), tailored filters are appealing
for measuring the low-mass end of the IMF. We describe here the design,
implementation, and characterization of a new near-IR filter tailored
for selection of brown dwarfs using ground-based telescopes. Our
approach maintains the efficiency of wide-field imaging, namely
measuring the properties of many objects in a single pointing, while
also providing spectral information. The results is a very high
confirmation rate for spectroscopic followup. Section~2 provides the
quantitative design of the filter, and Section~3 provides the
performance analysis. Section~4 presents our initial use of this filter
on the UH 88 inch telescope. Section~5 describes the photometric
selection and spectroscopic followup, with Section~6 discussing the confirmation rate of our survey. Finally, Section~7
gives our conclusions and discusses future directions.

\section{Design Of A Custom Near-IR Filter} \label{design}

We aimed to design a custom near-IR filter that photometrically selects
brown dwarfs in star-forming regions. We employed the following criteria
when considering possible filters:
\begin{enumerate}
\item In combination with broad-band photometry, our custom filter must be able to distinguish late-M and L dwarfs from background field stars \emph{independent of interstellar reddening}.  For the ages of star-forming regions and young clusters, the stellar/substellar boundary occurs at a spectral type of $\sim$M6 \citep[e.g.,][]{basri96, martin98, alves10}.
\item Our filter must lie in the wavelength range of 1.0--1.8~$\mu$m, a limitation set by the wavelength range of ULBCam on the UH 88 inch telescope.
\item To maximize observing efficiency, the filter should have as wide a bandwidth as possible.
\end {enumerate}

Of course, such a filter would also be suitable for selecting
  low-mass stars and brown dwarfs not in star-forming regions, where
  there is no extinction, though in this case traditional broad-band
  filters would generally offer superior performance.

\subsection{Synthetic Photometry} \label{sec:synphot}

To test the ability of various possible filter designs to discriminate between field stars and young brown dwarfs, we compiled a training sample of low-mass stars and brown dwarfs having high quality near-IR spectra.  We used:
 \begin{itemize}
 \item Our own SpeX near-IR spectra of optically classified young brown dwarfs 
in nearby star-forming regions (ages of $\sim$1--5~Myr) having spectral types ranging from M5 to 
M8 (Table \ref{tbl:spex_training}) .
\item Published spectra of young objects in IC~348 and Taurus (ages of $\sim$3 Myr) from \citet{muench07} and \citet{allers09}. 
\item Published near-IR spectra of young M and L dwarfs with {\sc VL-G} gravity classifications (estimated ages of $\sim$10~Myr) from \citet{allers13}.
\item Published M1--L9 dwarf spectral standards (expected to have typical solar neighborhood ages of a few Gyr) from \citet{cushing05} and \citet{kirkpatrick10}.
\end{itemize}


\begin{deluxetable}{lllllllllll}
\tablecaption{IRTF/SpeX Observing Log for Training Sample Brown Dwarfs in Star-Forming Regions\label{tbl:spex_training}}
\tabletypesize{\scriptsize}
\tablewidth{0pt}
\tablehead{
 \colhead{Object} &
 \colhead{SpT} &
 \colhead{Ref.} &
 \colhead{UT Date} &
 \colhead{Grat} &
 \colhead{Slit (\arcsec)} &
 \colhead{$R$} &
 \colhead{$\sec z$} &
 \colhead{$N_{\rm{exp}} \times t$ (s)} &
 \colhead{$T_{\rm{int}}$ (s)} &
 \colhead{$<S/N> (Y,J,H,K)$}
}

\startdata
2MASS~J03441558+3209218 & M7.5 & L16 &2001 Nov 6   &   ShortXD  &   0.5$\times$15 &   1200 &   1.17 &   14$\times$120.0 &   1680.0 &   18,  41,  43,  48 \\ 
2MASS~J03443103+3205460 & M5.5 & L16 &2001 Nov 6   &   ShortXD  &   0.5$\times$15 &    1200 &   1.13 &    8$\times$120.0 &    960.0 &   43,  94, 103, 109 \\ 
IfAHA 73 & M7.25 & L16 &2001 Nov 6  &   ShortXD  &   0.5$\times$15 &    1200 &   1.03 &    6$\times$120.0 &    720.0 &   41,  90,  93, 112 \\ 
2MASS J03443545+3208563 & M5.25 & L16 &2001 Nov 6   &   ShortXD  &   0.5$\times$15 &    1200 &   1.21 &    6$\times$120.0 &    720.0 &   55, 123, 149, 168 \\ 
2MASS J03443551+3208046 & M5.25 & L16 &2001 Nov 7   &  LowRes15  &   0.5$\times$15 &      120 &   1.19 &    6$\times$120.0 &    720.0 &  189, 273, 216, 166 \\ 
2MASS J03443595+3209243 & M5.25 & L16 &2001 Nov 7   &  LowRes15  &   0.5$\times$15 &      120 &   1.03 &    6$\times$120.0 &    720.0 &  155, 254, 200, 138 \\ 
2MASS J03443814+3210215 & M6 & L16 &2001 Nov 6   &   ShortXD  &   0.5$\times$15 &    1200 &   1.51 &   14$\times$120.0 &   1680.0 &   49, 105,  94, 109 \\ 
2MASS J03443886+3206364 & M6 & L16 &2001 Nov 6   &   ShortXD  &   0.5$\times$15 &    1200 &   1.07 &    6$\times$120.0 &    720.0 &   51, 102, 105, 112 \\ 
2MASS J03443920+3208136 & M8 & L16 &2001 Nov 7   &  LowRes15  &   0.5$\times$15 &      120 &   1.04 &    6$\times$120.0 &    720.0 &   78, 150, 117,  86 \\ 
2MASS J03443943+3210081 & M5 & L16 &2001 Nov 6   &   ShortXD  &   0.5$\times$15 &    1200 &   1.52 &    8$\times$120.0 &    960.0 &   62, 109, 109, 112 \\ 
$[$SS94$]$ V410 X-ray 6 & M5.5 & B02 &2002 Nov 25   &   ShortXD  &   0.5$\times$15 &   1200 &   1.01 &    8$\times$120.0 &    960.0 &  190, 256, 216, 195 \\
$[$BHS98$]$ MHO 5 & M6 & B02&2002 Nov 25   &   ShortXD  &   0.5$\times$15 &   1200 &   1.07 &    7$\times$120.0 &    840.0 &  115, 187, 163, 153 \\
$[$BHS98$]$ MHO 8 & M6 & B02 &2002 Nov 25   &   ShortXD  &   0.5$\times$15 &   1200 &   1.51 &    6$\times$120.0 &    720.0 &  157, 289, 245, 173 \\
$[$MDM2001$]$ CFHT-BD-Tau 2 & M7.5 & B02 &2001 Nov 6   &   ShortXD  &   0.5$\times$15 &    1200 &   1.69 &    6$\times$120.0 &    720.0 &   27,  75,  91, 116 \\
GM Tau & M5 & H14 &2002 Nov 25   &   ShortXD  &   0.5$\times$15 &   1200 &   1.71 &    6$\times$120.0 &    720.0 &   33, 107, 102, 102 \\
ISO-Oph 32 & M8 & W05 &2004 Jul 5   &   ShortXD  &   0.5$\times$15 &   1200 &   1.43 &    8$\times$180.0 &   1440.0 &  135, 260, 204, 205 \\
ISO-Oph 102 & M5.5 & W05 &2004 Jul 5   &   ShortXD  &   0.5$\times$15 &   1200 &   1.48 &    8$\times$180.0 &   1440.0 &  150, 280, 263, 319 \\
\enddata
\tablecomments{The last column gives the median S/N of the spectrum within the standard IR bandpasses}
\tablerefs{B02:~\citet{briceno02}; H14:~\citet{herczeg14}; L16:~\citet{luhman16}; W05:~\citet{wilking05}}

\end{deluxetable}

For each object in our training sample, we calculated synthetic photometry for the $J_{\rm{MKO}}$ and $H_{\rm{MKO}}$ filters \citep{tokunaga02}.   We then synthesized photometry for ideal (100\% transmission) filters centered at wavelengths of $1.0 < \lambda_{\rm{center}} < 1.8~\mu$m (in central wavelength steps of 0.01~$\mu$m) and filter bandwidths equal to 2\%, 4\%, 6\%, and 8\% of the central wavelength.  Thus, our design analysis considered 324 potential filters.

For each object, we calculated the flux through each filter as
\begin{equation}
\langle F_{\lambda, W} \rangle = \frac{\int \lambda F_\lambda(\lambda) S(\lambda) d\lambda}{\int \lambda S(\lambda) d\lambda}
\label{eq:fluxcalc}
\end{equation}
$S(\lambda)$ is the total system response including the atmospheric 
transmission\footnote{We used the atmospheric absorption spectrum for Maunakea
  from the Gemini Observatory webpages (\url{http://www.gemini.edu/sciops/telescopes-and-sites/observing-condition-constraints/ir-transmission-spectra}) for an airmass of 1.5 and precipitable water vapor content
  of 1.6~mm.}
 and filter response.  We did not include the 
throughput of the telescope and instrument nor the 
quantum efficiency of the detector, as our filter design 
could potentially be used in multiple near-IR imagers.  
Since we used photometric colors in subsequent analysis, we did not apply absolute flux calibrations to our synthetic photometry.  
To enable transformation between flux ratios and colors (in the Vega photometric
system), we also computed synthetic photometry for the Vega spectrum of
\citet{bohlin07}\footnote{\url{ftp://ftp.stsci.edu/cdbs/calspec/ascii/alpha_lyr_stis_005.ascii}} and assume zero color for Vega.

\subsection{Reddening-independent Index} \label{sec:qcalc}

Since we intended to use our custom filter to find young brown dwarfs in
star-forming regions, where interstellar extinction can be
significant, we wanted to be able to distinguish candidate members
from foreground and background field stars independent of reddening.  Following a similar method as \citet{najita00}, we defined a reddening-independent index
\begin{equation}
Q = J - W + e \times (H-W)
\label{eq:qcalc}
\end{equation}
$W$ refers to the magnitude of the object in our proposed filter, $J$ and $H$ are magnitudes in the standard broad-band filters, and $e$ is the ratio of reddening color excesses
\begin{equation}
e = \frac{A_J-A_W}{A_W-A_H}
\label{eq:ecalc}
\end{equation}

For each filter ($J, H$ and every potential custom filter), we calculated the selective
extinction, $A_\lambda$, as follows:  to represent the background population, first we reddened the spectrum of Gl~270 \citep{kirkpatrick10} by $A_V$=10 mag using the $R_V$=3.1 reddening law of \citet{fitzpatrick99}.   Next, we calculated the flux (Equation \ref{eq:fluxcalc}) through the filter for both the reddened and unreddened spectrum.  We then calculated the selective extinction for each filter as:
\begin{equation}
A_\lambda = -2.5~\log{(\frac{F_{\lambda, reddened}}{F_\lambda})} / A_V
\label{eq:alam}
\end{equation}
Using the selective extinction values for
$J, H$ and a potential custom filter, we calculated $e$ and determined $Q$,
the reddening insensitive index of each object in our training sample for each of our 324
potential filters.

The near-IR spectra of M and L dwarfs contain a wealth of features that are sensitive to effective temperature and gravity \citep[e.g.~see][and references therein]{allers13}.  Many of these features are alkali lines or molecular bandheads that are too narrow to be accurately quantified by a photometric filter.  Our custom filter is thus most sensitive to differences in continuum variations, in particular the deep and broad H$_2$O and FeH absorption features that shape the $J$ and $H$ band continuua of M and L dwarfs.  These broad H$_2$O and FeH continuum features vary monotonically across the late-M and L spectral range \citep[e.g.][]{cushing05, testi09} so the ability to cull contaminants from a photometric sample is tied to our custom filter's sensitivity to spectral type.  

\begin{figure}
\vskip 0.5in
\centerline{\includegraphics[angle=0, width=9cm]{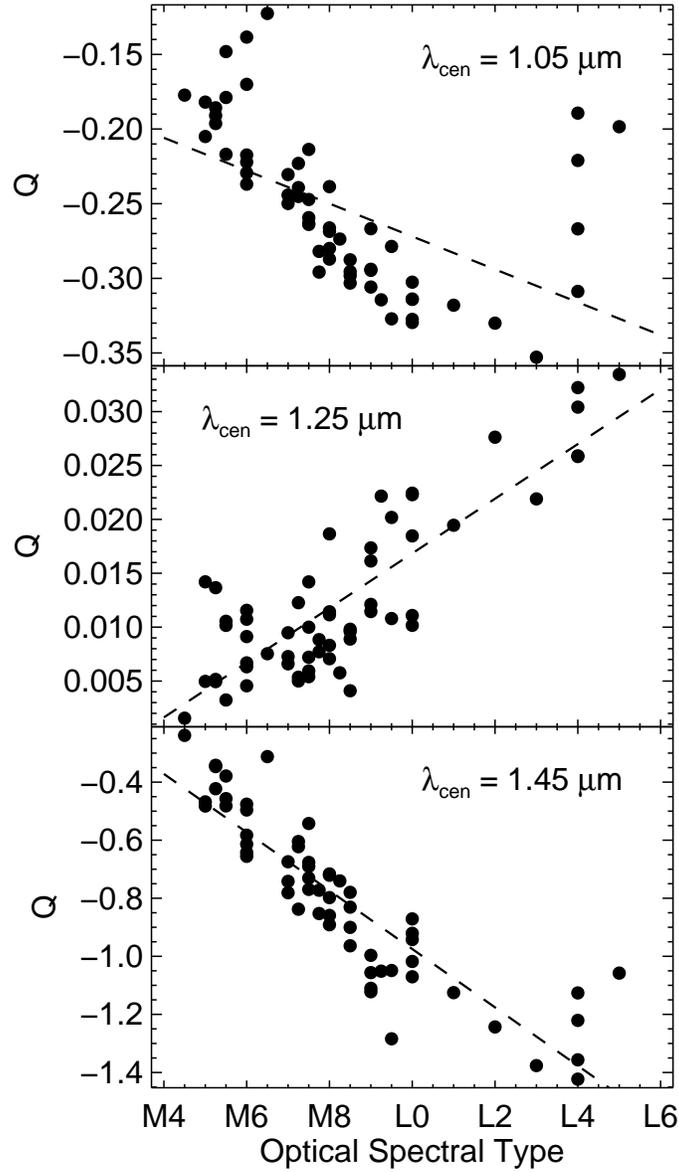}}
\vskip 2ex
\caption{\label{fig:q_spt_test} Reddening insensitive indices ($Q$) calculated from synthetic photometric (\S \ref{sec:synphot}) of our training sample for three potential filters.  $Q$ values are calculated as described in \S \ref{sec:qcalc} for filters with central wavelengths of 1.05 (top panel), 1.25 (middle panel), and 1.45~$\mu$m (bottom panel) and  bandwidths equal to 6\% of the central wavelength.  Dashed lines show the results of a linear fit to $Q$ vs. optical spectral type. }
\end{figure}

\begin{figure}
\vskip 0.5in
\centerline{\includegraphics[angle=0, width=12cm]{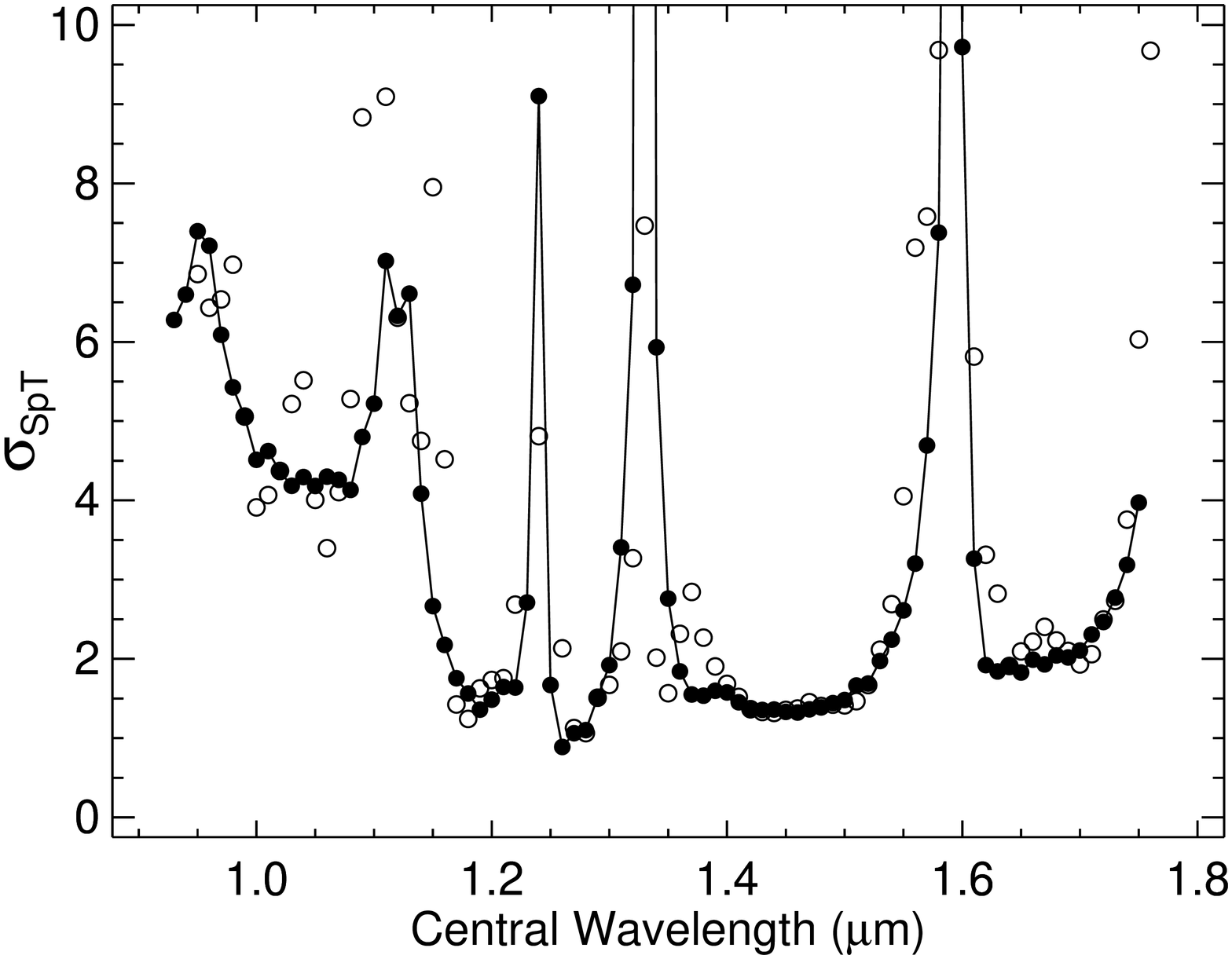}}
\vskip 2ex
\caption{\label{fig:spterr_wl_test} Spectral type uncertainty (in subtypes) vs.~filter central wavelength.  The spectral type uncertainty was determined from the scatter ($\sigma_Q$) about a linear fit of $Q$ vs. SpT for each filter and divided by the slope of the fit.  Solid circles (connected by a solid line) show $\sigma_{\mathrm{SpT}}$ for 6\% wide filters.  Open circles show $\sigma_{\mathrm{SpT}}$ for 2\% wide filters.  Based on this analysis, we decided to fabricate a 6\% wide filter centered at 1.45~$\mu$m.}
\end{figure}

We examined our custom filter's sensitivity to spectral type for our training sample.
Figure \ref{fig:q_spt_test} shows examples of $Q$ vs.~$SpT$ for three possible filters.  To determine the
best central wavelength for our filter, we calculated the scatter ($\sigma_Q$) of the $Q$ values about a linear fit of
$Q$ vs.~$SpT$ for objects in our training sample with SpT later than M6.
The uncertainty in SpT was determined as $\sigma_Q$ divided by the
slope of the linear fit ($dQ$/$d$(SpT)).  Figure
\ref{fig:spterr_wl_test} shows the uncertainty in SpT (as a result of
the scatter in $Q$) for filters with central wavelengths varying from
1.0 to 1.8~$\mu$m.  Based on the low SpT scatter, we chose to center our
filter at 1.45 $\mu$m.  

Having narrowed down the central wavelength of our filter to 
$\lambda_{\rm{center}}$=1.45~$\mu$m, we then determined the 
optimal filter bandwidth.  Ideally, the filter bandwidth should be as wide as possible to reduce integration time, 
without sacrificing sensitivity to SpT.  We 
tested filters of 2\%, 4\%, 6\% and 8\%, and found that the uncertainty 
and scatter in SpT changed by a negligible amount from 2\% to 6\% 
(Figure \ref{fig:spterr_wl_test}).  Thus, we
determined that the optimal custom filter for our purposes is a 6\%
wide filter centered at 1.45~$\mu$m. 

\section{Performance Analysis}

\begin{figure}
\vskip 0.5in
\centerline{\includegraphics[angle=0, width=12cm]{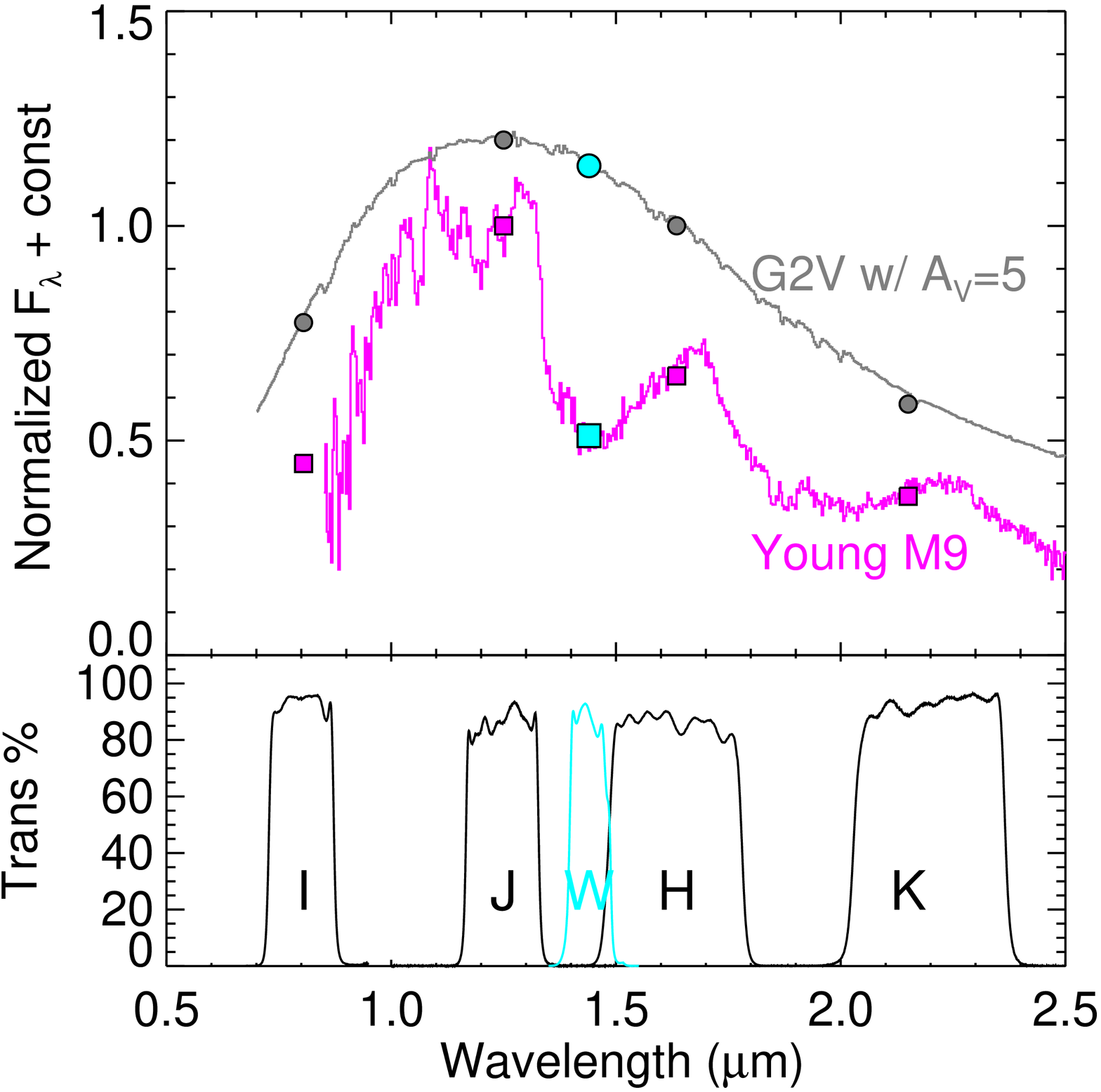}}
\vskip 2ex
\caption{\label{fig:filter_positions} Top: spectra and photometry for a G2V star \citep[HD~142093;][]{burgasser14} reddened by $A_V = 5 \units{mag}$ and a young M9 \citep[KPNO~12;][]{muench07}.  The broad-band $I,J,H,K$ colors of these two objects are nearly indistinguishable.  Bottom:  our custom $W$ filter breaks the degeneracy in broad-band colors for these two objects due to deep H$_2$O absorption in the spectra of ultracool dwarfs at the transmission wavelengths of our filter.}
\end{figure}

\begin{figure}
\vskip 0.5in
\centerline{\includegraphics[angle=0, width=12cm]{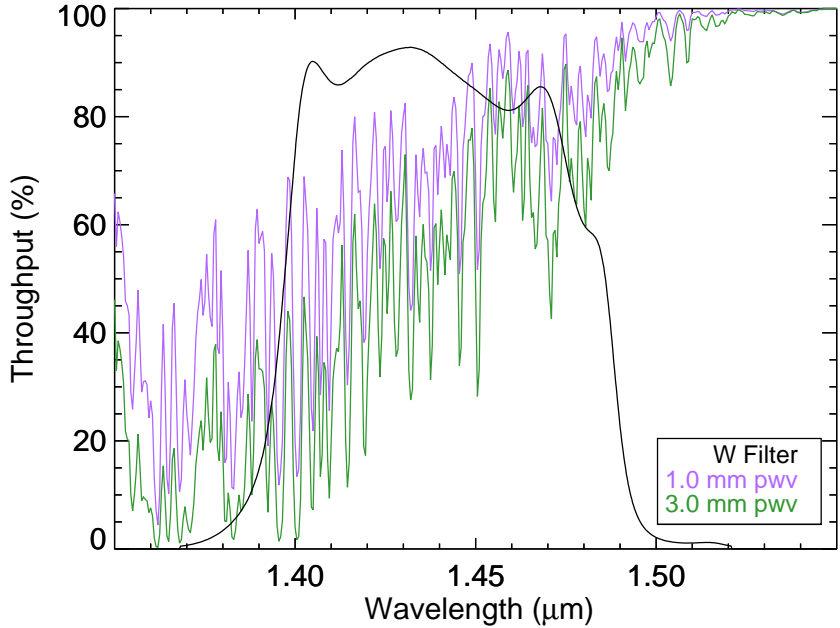}}
\vskip 2ex
\caption{\label{fig:filt_atmo} The throughput of our custom filter (black) compared to the sky transparency for precipitable water vapor contents of 1.0 and 3.0~mm and an airmass of 1.5.  The telluric absorption spectra calculated using the ATRAN software \citep{lord92} calculated for the conditions on Maunakea and available for download on the Gemini Observatory webpages (\url{http://www.gemini.edu/sciops/telescopes-and-sites/observing-condition-constraints/ir-transmission-spectra}).}
\end{figure}

\begin{figure}
\vskip 0.5in
\centerline{\includegraphics[angle=0, width=12cm]{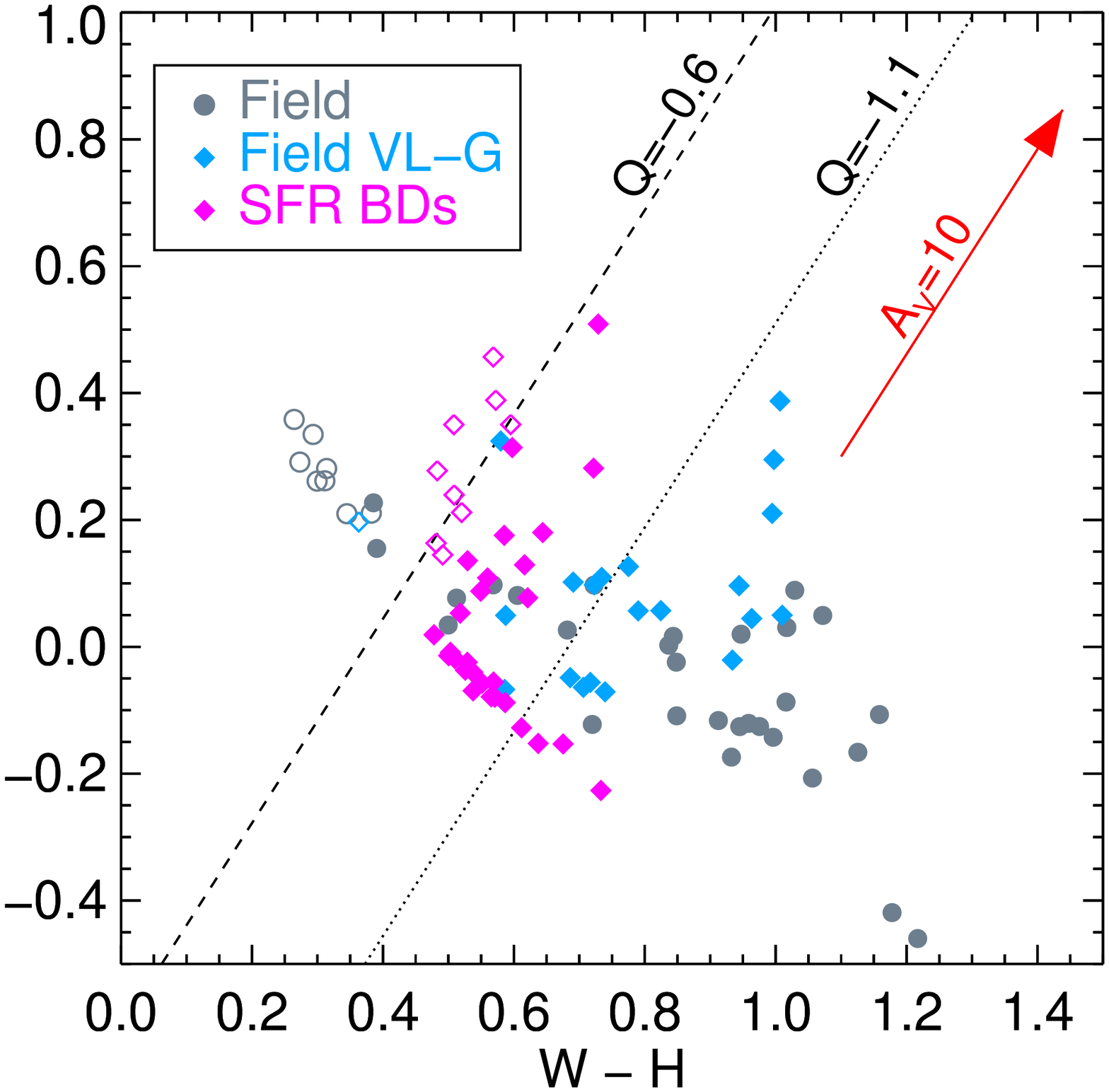}}
\vskip 2ex
\caption{\label{fig:color} $W-H$ and $J-W$ colors synthesized for our training sample (\S \ref{sec:synphot}).  The gray circles are synthetic photometry of M1--T5 field dwarf spectral standards \citep{cushing05,kirkpatrick10}.  The blue diamonds are young field brown dwarfs with gravity classifications of {\sc VL-G} \citep{allers13}.  The magenta diamonds are young objects in the Taurus, IC~348, and Ophiuchus star-forming regions \citep[Table \ref{tbl:spex_training};][]{muench07, allers09}.  A reddening vector for $A_V = 10$ is shown.  The dashed line denotes the colors of objects having a reddening-insensitive index of $-0.6$ and roughly indicates the substellar boundary for young objects.  The dotted line denotes the colors of objects having a reddening-insensitive index of $-1.1$, which corresponds to the M/L spectral type transition.  Solid symbols indicate objects with spectral types of M6 and later.  Open symbols indicate objects with spectral types earlier than M6.}
\end{figure}

\begin{figure}
\vskip 0.5in
\centerline{\includegraphics[angle=0, width=12cm]{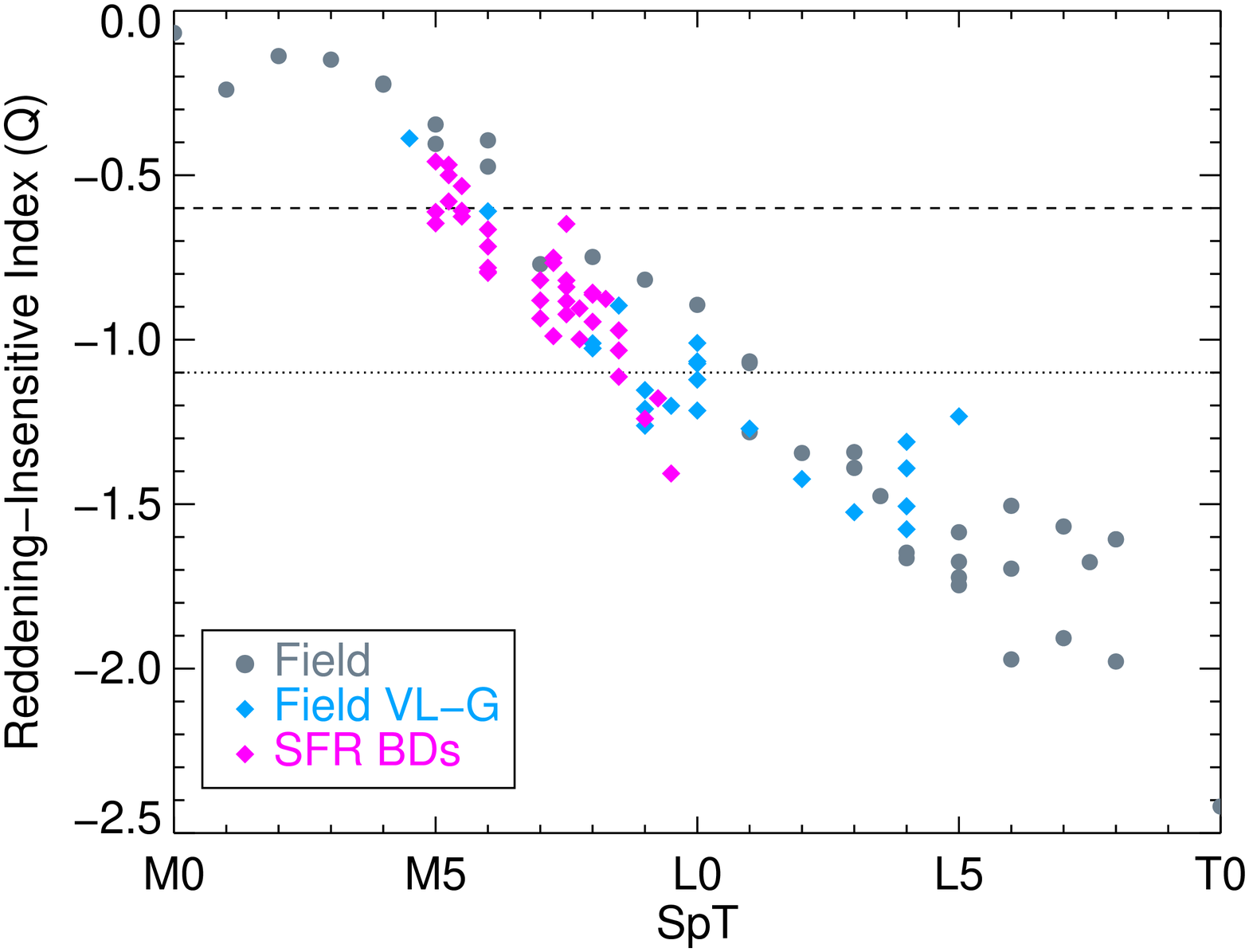}}
\vskip 2ex
\caption{\label{fig:spt_q} Relation between spectral type and the reddening-insensitive index, $Q$, calculated from synthesized $J , H,$ and $W$-band photometry of our training sample (\S \ref{sec:synphot}). The legend is the same as for Figure \ref{fig:color}. The dashed line denotes $Q=-0.6$, below which the $Q$ value indicates a spectral type later than M6, corresponding to the substellar boundary for young objects.  The dotted line denotes $Q=-1.1$, corresponding to the M/L spectral type transition.}
\end{figure}

Our filter was fabricated by Barr and Associates.
Figures \ref{fig:filter_positions} and \ref{fig:filt_atmo} show the transmission profile of a witness sample created at the same time as our filter and scanned while cooled to 50~K.  We use this cold witness scan in subsequent analysis, but also conducted a warm scan of our filter and found negligible differences in the filter throughput.  As shown in Figure \ref{fig:color}, our filter easily separates ultracool dwarfs from field stars using only $J, H,$ and $W$-band photometry, thanks to its sensitivity to deep H$_2$O absorption in the atmospheres of late-M and L dwarfs.  
Figure \ref{fig:spt_q} shows the reddening-insensitive index $Q$ (Equation \ref{eq:qcalc}) determined for our training sample as a function of spectral type.  Based on the scatter about a linear fit to SpT versus $Q$ for young M6--L5 type objects, our filter should be able to determine spectral type with a precision of 1.4 subtypes.  

\begin{figure}
\vskip 0.5in
\centerline{\includegraphics[angle=0, width=12cm]{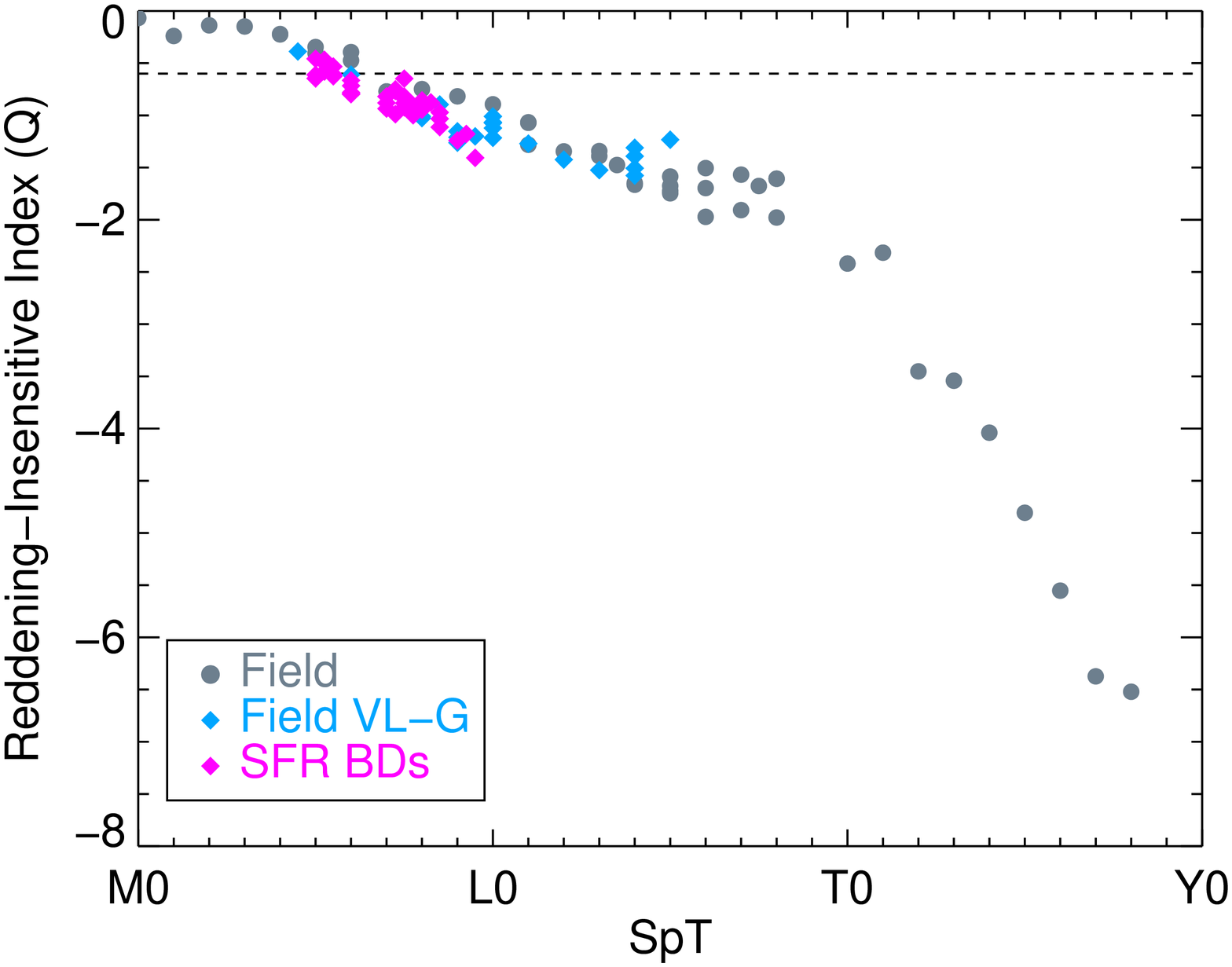}}
\vskip 2ex
\caption{\label{fig:spt_q_extend} Same as Figure \ref{fig:spt_q}, but showing an extension to T dwarfs using spectral standards from \citet{kirkpatrick10}.}
\end{figure}

The majority of objects in our training sample have spectral types of M5-L5, for which $Q$ shows a linear dependence on spectral type.  Cooler T dwarfs also have significant H$_2$O absorption, but their methane-dominated spectra do not follow the same linear trend seen for M5--L5 objects, and the absorption at 1.45~$\mu$m is so deep that photometric sensitivity can be a limiting factor for detection.  Nonetheless, our custom filter can distinguish T dwarfs from background stars (Figure \ref{fig:spt_q_extend}), though broad band and methane filters may be better suited to their detection and characterization.

\subsection{\wfilt\ Filter Fluxes and Wavelengths for Vega}
Using the spectrum of \citet{bohlin07}, we calculate the flux density ($F_\lambda$), isophotal wavelength ($\lambda_{\rm{iso}}$), and effective wavelength ($\lambda_{\rm{eff}}$) for Vega in the \wfilt\ filter.  We first scale the Vega spectrum to $3.46 \times 10^{-9}$~W~m$^{-2}$~$\mu$m$^{-1}$ at a wavelength of 0.5556~$\mu$m \citep{megessier95}.  For our filter, we calculate a flux density of $1.79 \times 10^{-9}$~W~m$^{-2}~\mu$m$^{-1}$ and isophotal and effective wavelengths of 1.445~$\mu$m for Vega, using a system response that includes our filter profile and atmospheric transmission for 1.6~mm of precipitable water vapor and an airmass of 1.5.  The \wfilt-band flux density of Vega in frequency units\footnote{Calculated as $\langle F_{\nu, W} \rangle = \frac{\int F_\nu(\nu) S(\nu)/\nu d\nu}{\int S(\nu)/\nu d\nu}$ from the Vega spectrum, scaled to have $F_{\nu} = 3630$~Jy in $V$-band.} is $F_\nu = 1180 \units{Jy}$.  Using Equation 9 from \citet{tokunaga05}, we calculate an AB magnitude of $\wfilt_{\mathrm{AB}} =1.25 \units{mag}$ for Vega.


\subsection{Comparison of Integration Times}
To determine the necessary integration times for our custom filter, we computed the total flux from Vega that makes it onto the detector ($\int \lambda F_\lambda(\lambda) S(\lambda) d\lambda$), assuming that losses due to reflections and detector quantum efficiency are wavelength-independent and using a telluric transmission spectrum for 1.6~mm of precipitable water vapor at an airmass of 1.5.  For Vega, the detected flux in the \wfilt-band is 0.26 times that in the $H$-band.  The shorter wavelength and narrower bandwidth of the \wfilt\ filter compared to $H$, however, means that the \wfilt\ filter has 7.1 times lower sky background.\footnote{We calculated the sky background using the ATRAN models \citep{lord92} for a pwv of 1.6~mm and an airmass of 1.5.  The sky background spectrum also includes the contribution of a 273~K blackbody as described on the Gemini webpages (\url{http://www.gemini.edu/sciops/telescopes-and-sites/observing-condition-constraints/ir-background-spectra}).}  We calculated the relative noise levels assuming that the photometric noise is dominated by Poisson noise of the sky background (i.e.~neglecting readnoise and moonlight), and that the seeing is identical in \wfilt-band and $H$-band.  We find that to reach the same Vega-magnitude with the same signal-to-noise ratio (S/N), observations using the \wfilt\ filter require $\approx$2 times the integration time required for $H$-band.  For young late-M and L type objects, the intrinsic \wfilt-band magnitude is fainter than in $H$-band (Figure \ref{fig:color}), so even deeper \wfilt-band observations are needed to achieve the same S/N as $H$-band observations.

\subsection{The Effects of Precipitable Water Vapor}\label{sec:pwv}

The deep H$_2$O absorption features that allow our filter to identify M and L-type objects coincide with deep telluric absorption features.  Thus, the throughput of the filter will depend on airmass and the atmospheric precipitable water vapor (pwv) content.  Figure \ref{fig:filt_atmo} shows the throughput of the \wfilt\ filter as well as telluric transmission for precipitable water vapor contents of 1.0 and 3.0~mm (at an airmass of 1.5) which correspond to the 20th and 80th percentile of conditions on Maunakea.\footnote{\url{http://www.gemini.edu/sciops/telescopes-and-sites/observing-condition-constraints/mk-water-vapour-statistics}}
The median precipitable water vapor for Maunakea is 1.6~mm, which is what we use in our determination of $e$ (Equation 3).   For the spectrum of Vega, a change in pwv from 1~mm to 3~mm results in a 26\% reduction in total system response ($S(\lambda)$ in Equation 1).   Similarly, under median pwv conditions (1.6~mm), a change in airmass from 1 to 2 results in a 17\% reduction in total system response.  Thus, changes in atmospheric pwv and airmass affect the zero-point of the \wfilt\ filter, so photometric calibration should not be done using non-contemporaneous photometric standard fields.

To see if pwv will affect our calculation of $Q$, we can look at the change in effective wavelength.
For a pwv of 1~mm (at an airmass of 1.5), the effective wavelength of the \wfilt\ filter for Vega is 1.4436~$\mu$m.  For a pwv of 3~mm, the effective wavelength shifts very slightly redward, to 1.4470~$\mu$m.  The calculated flux density for Vega also has a very small change with pwv.  A change in pwv from 1 to 3~mm changes Vega's calculated flux density by only 0.8\%.  Thus, though preciptable water vapor changes the throughput of the filter, if care is taken in photometric calibration (\S \ref{sec:phot}), pwv has a negligible effect on the photometric colors and calculated $Q$ values.

\subsection{Selective Extinction} \label{sec:ecalc}

The effects of interstellar extinction are wavelength-dependent.  The selective extinction for a given filter, $A_\lambda$, is dependent on the wavelength of filter as well as the spectral shape of the light being attenuated by the dust.  The selective extinction ratio, $A_\lambda/A_V$, thus depends on the interstellar reddening law (i.e. properties of the attenuating dust), the amount of extinction, and the spectral energy distribution of the object being attenuated \citep[e.g.~see][]{fiorucci03}.  We calculate the selective extinction for spectra of Vega \citep{bohlin07}, an M0 \citep[Gl207,][]{kirkpatrick10} and a young L0 \citep[the L0 VL-G standard, 2M2213-12 from ][]{allers13}.  Using Equation \ref{eq:alam}, we calculate the selective extinctions $A_J, A_H$, and $A_W$ using the reddening law of \citet{fitzpatrick99} for $R_V$=3.1 and 5.0 and $A_V$ = 1, 5, 10, and 20~mag.  We find that the values of selective extinction change very little.  For example, at all values of $A_V$ and $R_V$ the value of $A_W/A_V$ ranges only from 0.197 to 0.203 for all three spectra.  


\begin{deluxetable}{crrrrrr}
\tablecaption{Selective Extinction ($e$) Ratios}
\tabletypesize{\footnotesize}
\tablewidth{0pt}
\tablehead{
  \colhead{SpT\tablenotemark{a}} & \colhead{$A_V$} & \colhead{$R_V$} & \colhead{Airmass} &
  \colhead{pwv} & \colhead{$e_{\mathrm{MKO}}$} \\
  \colhead{} & \colhead{(mag)} & \colhead{} &
  \colhead{} & \colhead{(mm)} & \colhead{} }
\startdata
M0 & 10.0 & 3.1 & 1.5 & 1.6 & {\bf 1.61}\tablenotemark{a} \\
\hline
M0 & 10.0 & 5.0 & 1.5 & 1.6 & 1.74 \\ 
M0 & 1.0   & 3.1 & 1.5 & 1.6 & 1.67 \\ 
M0 & 20.0   & 3.1 & 1.5 & 1.6 & 1.55 \\ 
M0 & 10.0   & 3.1 & 1.0 & 1.0 & 1.58 \\ 
M0 & 10.0   & 3.1 & 2.0 & 3.0 & 1.67 \\ 
A0 & 10.0 & 3.1 & 1.5 & 1.6 & 1.74 \\ 
L0  & 10.0 & 3.1 & 1.5 & 1.6 & 1.56 \\ 
\enddata
\tablenotetext{a}{The value of $e$ adopted for our data reduction and analysis.}
\label{tbl:e}
\end{deluxetable}

Though the values for selective extinction vary little, the values for $e$ (Equation \ref{eq:ecalc}) have a much larger variation (1.46-1.93). Table \ref{tbl:e} summarizes the variation in $e$ with SpT, $A_V$, $R_V$, airmass, and pwv.  Since the goal is to distinguish between background stars and ultracool dwarfs, choosing a value of $e$ appropriate for the reddened background population ensures that fewer contaminants are selected by our filter.  The median $A_V$ for the regions of Perseus and Ophiuchus in our survey is 4 mag.  However, the choice of $e$ more strongly affects calculated $Q$ values (Equation \ref{eq:qcalc}) for highly reddened objects.
Thus, for calculating $Q$ from \wfilt, $J_\mathrm{MKO}$ and $H_\mathrm{MKO}$, we chose to adopt $e_\mathrm{MKO} = 1.61$ the value calculated for an M0 star reddened by $A_V = 10 \units{mag}$ and $R_V = 3.1$ and observed at an airmass of 1.5 and pwv content of 1.6.  Because \emph{Two Micron All Sky Survey} photometry is readily available, we also calculate $e_\mathrm{2MASS}=1.52$ using the 2MASS filter profiles, so that $W$ filter photometry can also be combined with $J_\mathrm{2MASS}$ and $H_\mathrm{2MASS}$.

We tested how our choice of $e$ would affect our selection of ultracool dwarf candidates by examining how our calculated $Q$ value (Equation \ref{eq:qcalc}) and corresponding SpT (from the linear relation shown in Figure \ref{fig:spt_q}) for a young L0 are affected by variations in reddening and observing conditions.
For moderate reddening ($A_V < 10 \units{mag}$), we find that our adoption of $e_{\mathrm{MKO}} = 1.61$ leads to variations in the determined spectral type of a reddened young L0 of $<$0.5 subtypes, well below the scatter in the SpT versus $Q$ relation in Figure \ref{fig:spt_q}.  

Our choice of $e$ can cause reddened background stars to have artificially lower $Q$ values.  For background objects with spectral types of M0, moderate reddening ($A_V < 10 \units{mag}$) causes a negligible change in calculated $Q$ values, as expected for a reddening-insensitive index.  For highly reddened objects, however, $Q$ has some sensitivity to extinction.  If reddened by $A_V = 30 \units{mag}$, an M0 star has a calculated $Q$ value that is 0.2 lower than its $Q$ for moderate extinctions.  This indicates that care should be taken when applying $Q$ selection criteria to highly reddened objects.   

\section{Observations and Data Reduction}

We conducted a pilot survey by imaging two star-forming molecular clouds in broad-band $J$ and $H$ filters, as well as our custom \wfilt\ filter.  We chose to image the Ophiuchus and Perseus star-forming regions, as they are relatively nearby ($\sim$140~pc and $\sim$300~pc, respectively), which allowed us to efficiently verify candidates spectroscopically (\S \ref{sec:spexobs}).  In Ophiuchus, our survey covered $\sim$1.6 square degrees and included the L1688 and L1689 regions, as well as a region to the west of L1688 surveyed in \citet{allers06}.  The ages for sources associated with the Ophiuchus region are estimated to be between $\approx$0.3~Myr for embedded sources in L1688 region to $\approx$10~Myr for objects in the surface population \citep{wilking08}.  In Perseus, our survey covered $\approx$1.3 square degrees and included the IC~348 and NGC~1333 clusters, with ages of $\sim$2 and $\sim$1~Myr, respectively \citep{young15}.

\subsection{Broad-band J and H Photometry}

Broad-band $J$ and $H$ photometry was obtained using UKIRT/WFCAM (Project U/07A/H30), CFHT/WIRCAM (Programs 08AH14 and 07BH20) and Blanco/ISPI (Program 2010A-0326).

The data taken with the UKIRT Wide Field Camera \citep[WFCAM;][]{casali07} use a photometric system described in \citet{hewett06}, which includes $J$ and $H$-band filters meeting the Maunakea Observatories \citep[MKO;][]{simons02, tokunaga02} near-IR filter specifications.  The pipeline processing and science archive are described in \citet{irwin04} and \citet{hambly08}, respectively.   We used the pipeline-processed photometry tables from stacked images.  
  
For $J$ and $H$-band data taken with CFHT/WIRCAM, we used pre-processed data from 'I'iwi, the CFHT facility pipeline, which includes detrending and sky subtraction.  For data taken with Blanco/ISPI, we used custom IDL programs \citep{allers06} to flat field and sky subtract the data.  We determined astrometry for our WIRCAM and ISPI fields using astromatic's SCAMP program \citep{scamp} and then combined data for each pointing using SWARP \citep{swarp}.  We measured photometry using Source Extractor \citep{sextractor}.   We determine the photometric calibration for our data using 2MASS \citep{skrutskie06}.  We first convert the 2MASS photometry for our regions to the MKO system using the conversions of \citet{leggett06}.  In addition, we included $J$ and $H$ band photometry of Ophiuchus from \citet{allers06}.  We then combine all of our broad-band $J$ and $H$ photometry into a master catalog, using a weighted mean to combine photometry for objects detected in more than one field.  

\subsection{\wfilt\ Photometry} \label{sec:phot}

Observations with the \wfilt\ filter were obtained using the ULBCam
instrument on the UH 88'' Telescope on Maunakea.   ULBCam contained a 4096~$\times$~4096 pixel detector array, configured as a mosaic of four 2048~$\times$~2048 pixel prototype Hawaii2-RG detectors.  The array had a 17\arcmin~$\times$~17\arcmin\ field of view with a pixel scale of 0\farcs25.  Four amplifiers were used per detector, with each amplifier controlling 512~$\times$~2048 pixels.

ULBCam stored every read of the array.
We employed a technique whereby we reset the array, took an initial read, read
out the array again after 10~s and had a final read of the
array 210~s later.  We then moved to the
next position in our dither sequence, and took another set of three exposures.  For a typical field, we used a 13-point dither pattern with offsets of 0\farcm5--1\farcm0 from the initial position.  A typical observing sequence resulted in 45~minutes of total integration time.  The details of our observations are available in Table \ref{tbl:wobs}.

Data reduction for ULBCam data is complicated by many array
imperfections.  In particular, the array suffered from severe image
persistence, an occasionally dead or extremely noisy amplifier, and for our 2008 data, a
dead detector.  To account for these array imperfections in our data reduction, we construct a master flat field by mean-combining several dome and twilight flats taken during each observing campaign for which the instrument was contiguously installed on the telescope.  We also constructed a flat field error frame as the standard deviation of the mean.  A bad pixel mask was created by flagging dead pixels and pixels with a flat field uncertainty greater than 10\%.

We use custom IDL scripts to create a double-pass sky frame.  We first flag saturated pixels in the final read and then subtract the first read from the final read.  To determine the uncertainty due to flat field error ($\sigma_{\rm{flat}}$), we multiply each frame by the flat field error frame.  We then split each frame (including saturation flag, flat field uncertainty, bad pixel mask and read-subtracted frame) into individual images (of 512$\times$2048 pixels) for each amplifier.  We median-zero each image within a set, and create a median-combined initial sky frame which is then subtracted from each image.  Using the SExtractor program \citep{sextractor}, we identify stars in each image and save a background rms map ($\sigma_{\rm{bkg}}$).  We then create a second-pass sky
after masking out the detected stars, and subtract this sky from the images.  For each image, we also construct a weight map, where the weight for each pixel is: $w = \frac{bad pixel mask}{\sigma_{bkg}^2 + \sigma_{flat}^2}$.   Some frames showed significant smearing of point sources, indicative of slight motion of the telescope.  These frames were used in the determination of the sky frame but were discarded from the remaining steps in our data reduction process, and are not included in the total integration times listed in Table \ref{tbl:wobs}.

To create an astrometric solution for each image, we first re-combine the images from each amplifier to create a multi-extension FITS file where each extension corresponds to an individual amplifier.  For each observing semester (2007A, 2008A, and 2008B), we create a global astrometric header and use the SCAMP program \citep{scamp} to determine the astrometric solution of our images by matching to 2MASS $H$-band detections.  The median rms in position for bright detections (S/N~$>~100$) relative to positions in the 2MASS catalog was 0\farcs11. 

\begin{figure}
\vskip 0.5in
\centerline{\includegraphics[angle=0, width=12cm]{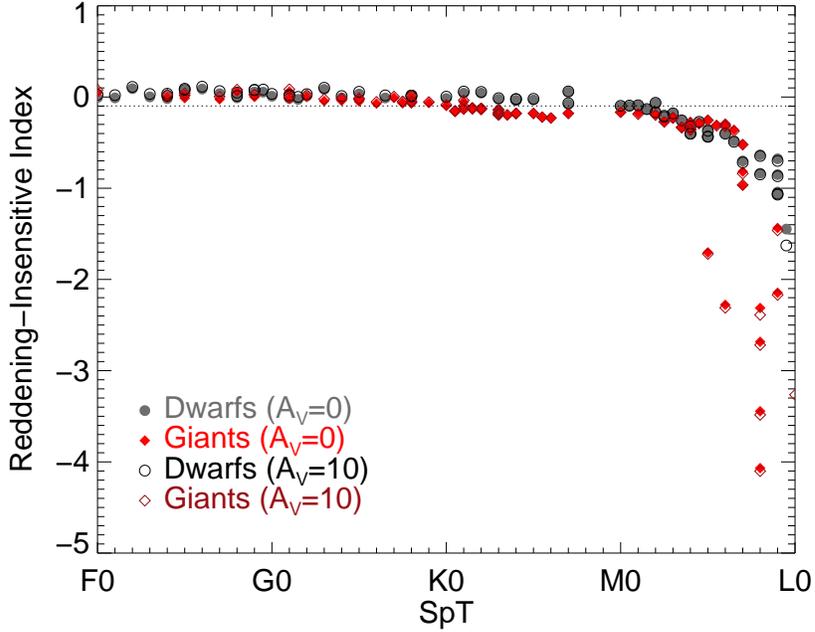}}
\vskip 2ex
\caption{\label{fig:spt_q_stars} Relation between spectral type and the reddening-insensitive index, $Q$, calculated from synthesized $J , H,$ and $W$-band photometry from the SpeX Spectral Library \citep{cushing05,rayner09}.  The population of background stars is expected to be dominated by late-K and early M dwarfs.  The dotted line denotes $Q$=-0.098, which we use to determine the flux zero-point for each of our $W$-filter images.}
\end{figure}

The H$_2$O absorption feature on which we centered our custom filter coincides with a telluric absorption feature.  Thus, the throughput of the filter and the photometric zero-point depend on airmass and the atmospheric precipitable water vapor content (see Section \ref{sec:pwv}).   
To determine the \wfilt\ magnitudes of our detected objects, we
employ the expected $Q$~values of background stars, which are roughly constant for spectral types earlier than M3 (Figure \ref{fig:spt_q_stars}).
Galactic population synthesis models \citep{robin03, trilegal} predict that the background population in our images will be dominated by late-K and early-M dwarfs.   

So we determine the photometric zero-point for each \wfilt\ image such that
the median measured $Q$ agrees with the median predicted $Q$ of the synthesized background ($-0.098$).   
Our photometric calibration uncertainties (based on the standard deviation of the mean of measured $Q$ values within an individual frame) range from 0.01~mag for the typical, low extinction regions with many detected stars to 0.05~mag for regions with high extinction and few detected stars.

\begin{figure}
\vskip 0.5in
\centerline{\includegraphics[angle=0, width=12cm]{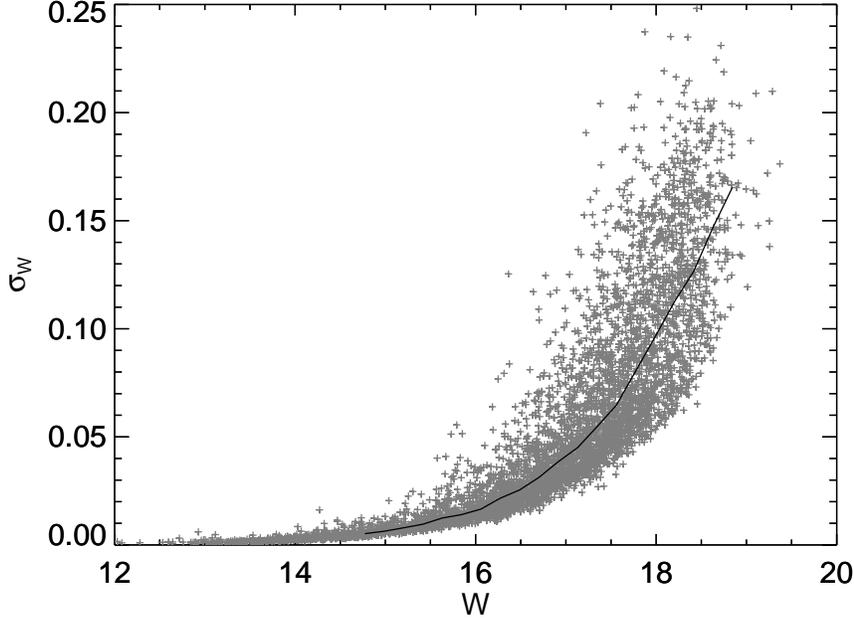}}
\vskip 2ex
\caption{\label{fig:wphot} \wfilt\ photometry for IC~348 fields observed on 2018 November 6.  The gray points show the final \wfilt\ photometric magnitudes and uncertainties.  The solid line shows the median photometric uncertainty as a function of magnitude.}
\end{figure}

Because of the array imperfections present in ULBCam data, we did not stack our images, but rather extracted photometry on individual reduced frames using SExtractor.  We then combined photometry for individual objects across all of our images using a robust weighted mean.  Figure \ref{fig:wphot} shows the \wfilt\ magnitudes and uncertainties for sources toward IC~348 observed on 2018 November 6.

\section{SpeX Followup of Brown Dwarf Candidates} \label{sec:spexobs}

\begin{figure}
\vskip 0.5in
\centerline{\includegraphics[angle=0, width=12cm]{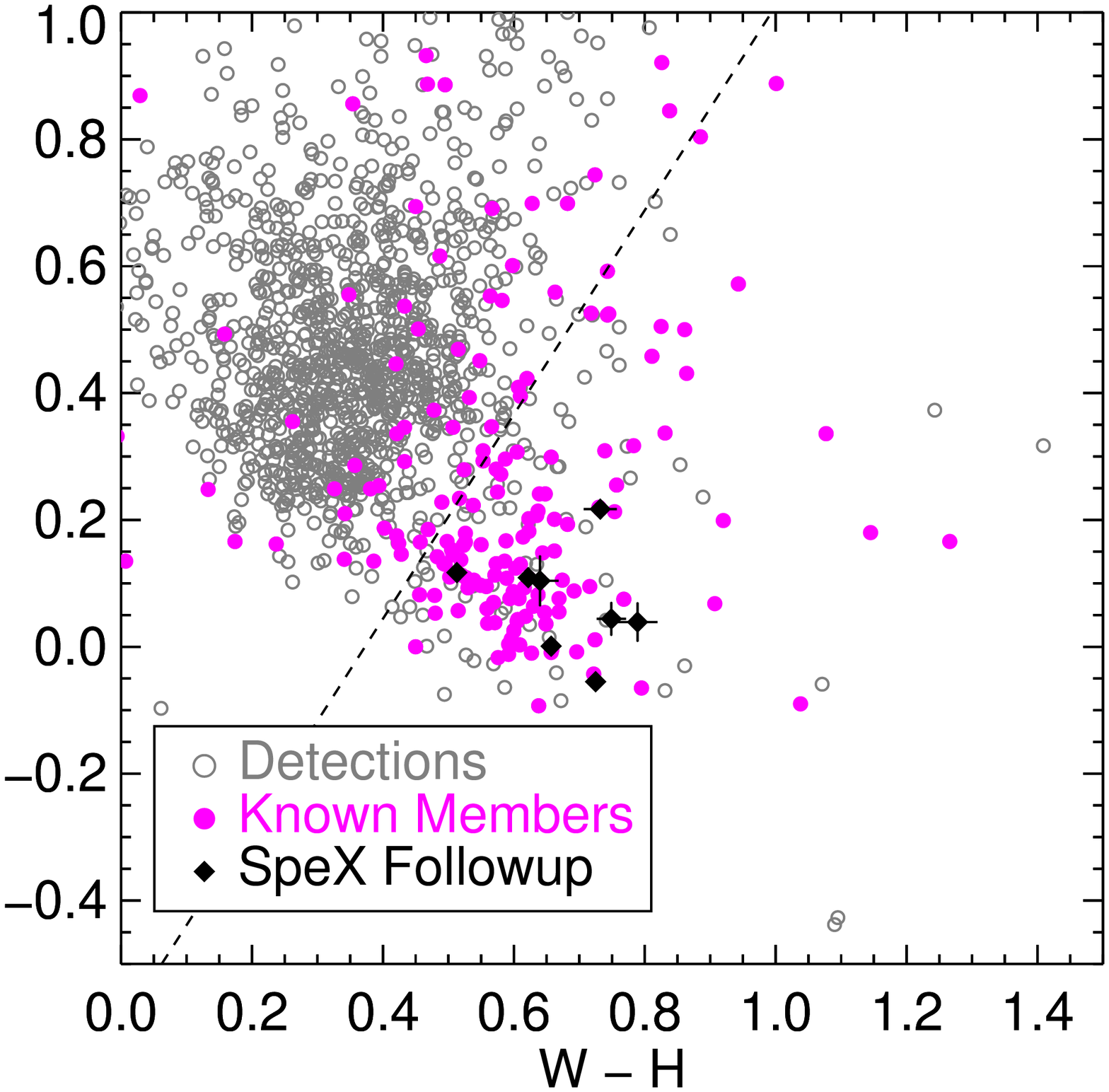}}
\vskip 2ex
\caption{\label{fig:obscolor} Color-color plot showing our selection criteria for IC~348 fields observed with the \wfilt\ on 2018 November 6.  The gray circles show all of the detections in the final photometric catalog.  The magenta circles show our photometry for IC~348 members compiled by \citet{luhman16}.  The dashed line indicates a line of constant $Q=-0.6$.  We selected candidates for spectroscopic followup that had $Q$ values $>3 \sigma$ below $-$0.6, which corresponds to objects which fall $>3\sigma$ to the right of the dashed line.  We obtained spectroscopic followup of 8 of the candidates (black diamonds), all of which were confirmed.}
\end{figure}

\begin{figure}
\vskip 0.5in
\centerline{\includegraphics[angle=0, width=12cm]{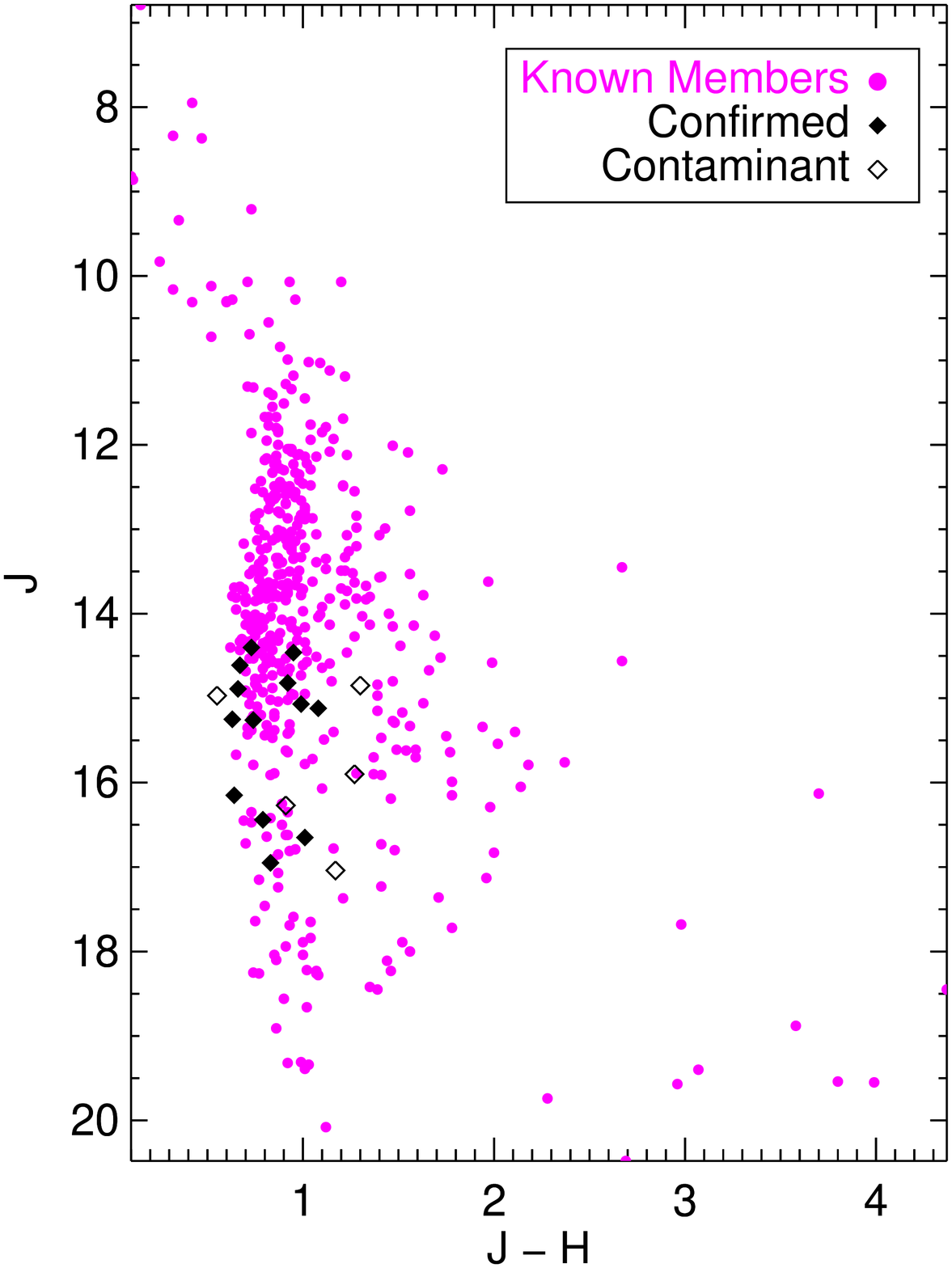}}
\vskip 2ex
\caption{\label{fig:colormag} Color-magnitude plot comparing our candidates with spectroscopic followup to known members of IC~348.  The magenta circles show photometry for IC~348 members compiled by \citet{luhman16}.  We obtained spectroscopic followup of 18 of the candidates (black diamonds), 13 of which were confirmed as likely IC~138 members (solid diamonds).}
\end{figure}

We selected our list of candidates \emph{using their $Q$-values alone}.
Although ancillary photometry (optical, Spitzer, and WISE)
were available, we chose to rely solely on the combination of $J, H,$ and
\wfilt\ photometry to test the efficacy of our custom filter
(Table~\ref{tbl:candphot}). We selected objects having $Q < -0.6$,
which should correspond to objects with spectral types later than M6.
Figure \ref{fig:obscolor} shows the criteria we used to identify
  candidates on the $J-W$ versus $W-H$ color plane.  Figure \ref{fig:colormag} compares the broad band colors and magnitudes of our candidates to known members.  Overall, our candidates do not have unusual broad band colors compared to previously known members.

We obtained near-IR spectroscopic followup for 60 of our candidate brown dwarfs using the SpeX spectrograph \citep{rayner03} on the NASA Infrared Telescope Facility.  The majority of our spectra were obtained using the low-resolution prism mode, which covers the full 0.8--2.5~$\mu$m wavelength range at low resolution ($R\sim75-200$).  For a few brighter candidates, we used the higher resolution cross-dispersed observing mode ($R\sim750-2000$). Table \ref{tbl:spex_observing} provides the details of our observations.   We observed A0 stars as telluric standards adjacent to our observations of our candidates, and selected the particular standard to provide a good match in sky position (typically within 10\degr) and airmass (typically within 0.1).  For IC~348 and NGC~1333 candidates, we used either HD~22859 or HD~24000 as telluric standards.  For Ophiuchus candidates, we used either HD~149827, HD~151787, or HD~153068 as telluric standards.
We used SpeXtool version 3.0 \citep{cushing04} to reduce our spectra,
including correction for telluric absorption following the method
described in \citet{vacca03}.

\begin{turnpage}
\begin{figure}
\includegraphics[width=3.7in]{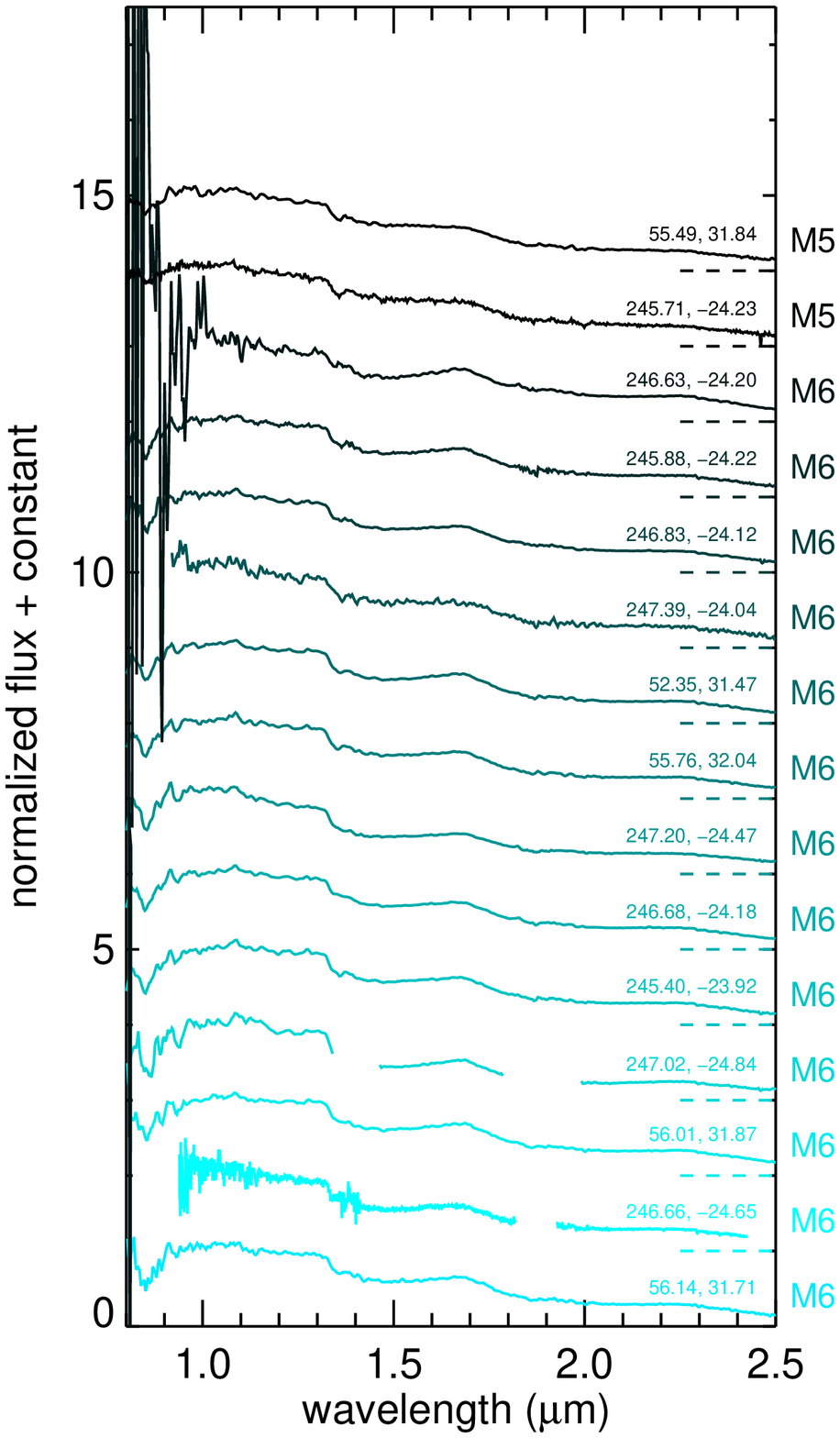}
\hskip -0.7in
\includegraphics[width=3.7in]{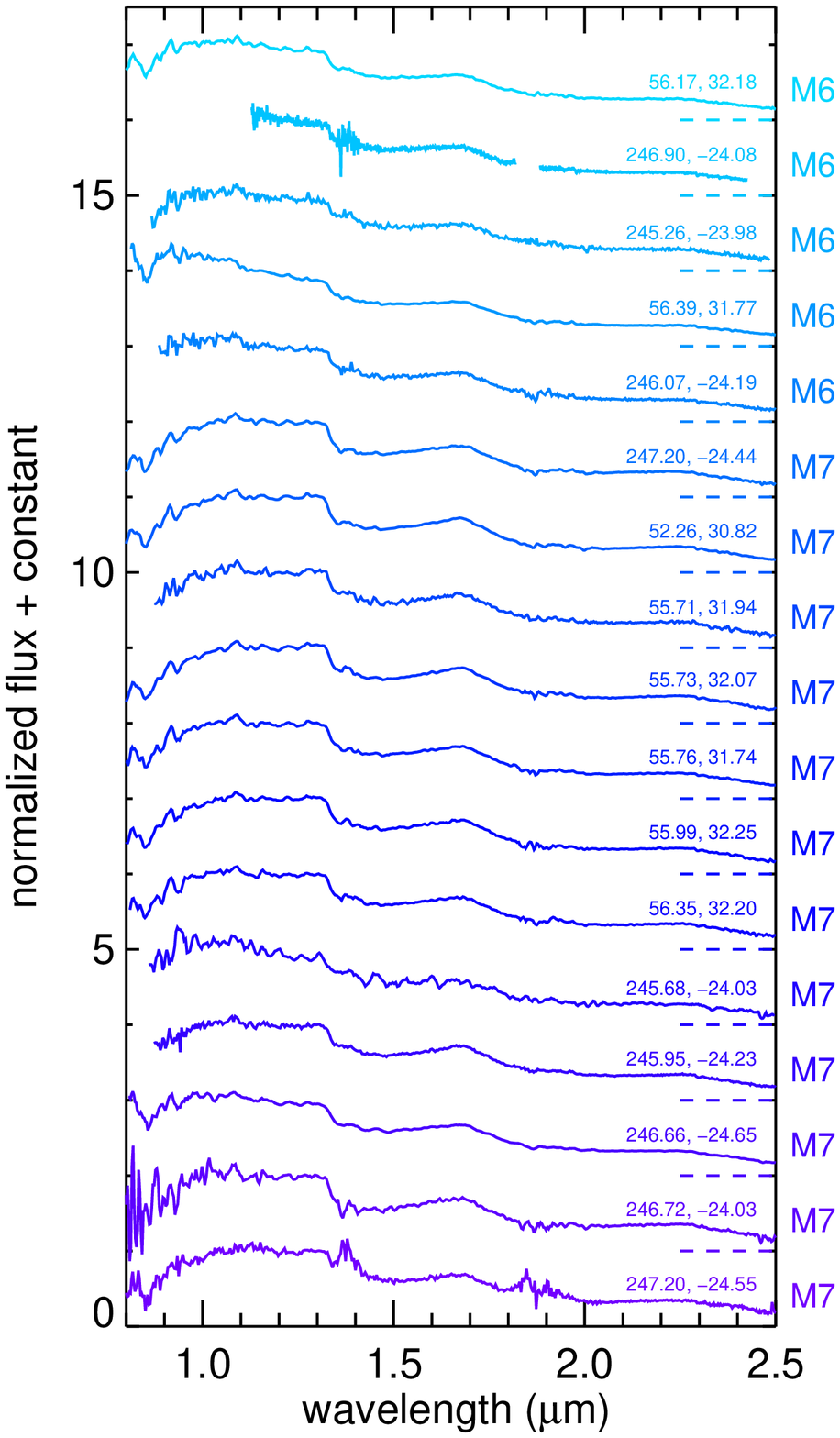}
\hskip -0.7in
\includegraphics[width=3.7in]{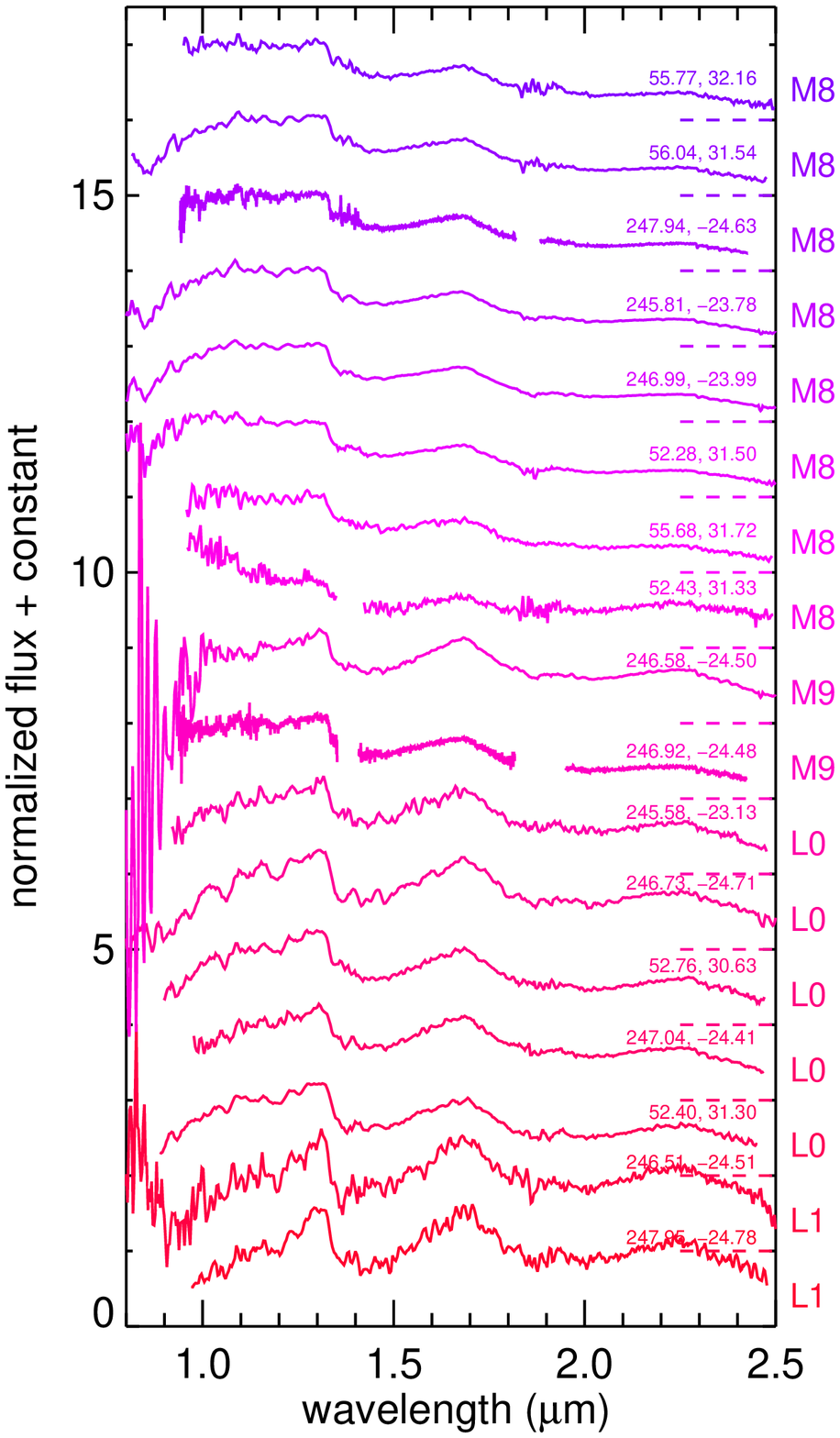}
\caption{\normalsize De-reddened spectra of candidate members with spectral types of M5 and later.  Spectra have been de-reddened using the reddening law of \citet{fitzpatrick99} using the estimated $A_V$ values in Table \ref{tbl:spts}.  Spectra are labeled with the decimal coordinates (in degrees) of the candidates. \label{fig:spectra}}
\end{figure}
\end{turnpage}

We classify our spectra using a method similar to that described in \citet{allers13}, which includes a gravity-independent determination of near-IR spectral type followed by an index-based gravity classification.  Unlike the \citet{allers13} sample, our targets are likely to be attenuated by dust from the star-forming region.  Thus, we modified the \citet{allers13} classification method as follows:
\begin{itemize}
\item{For visual comparison to spectral type standards, we use the VL-G spectral standards presented in \citet{allers13}.} 
\item{While comparing to spectral standards, we first de-redden the spectrum of our candidate so that its synthetic $J-K$ color matches the spectral standard.}
\item{To determine an estimated $A_V$ for our candidates, we use the mean of $A_V$ values needed to match spectral standards that have spectral types within one subtype (the uncertainty of the \citet{allers13} classification) of the best-matching standard.}
\item{We then de-redden the candidate's 
spectrum by our $A_V$ estimate, and follow the \citet{allers13} method for determining the overall spectral type and gravity classification (using our earlier determination of visual spectral types).}
\end{itemize}

The spectral types and gravity determinations for our spectra are presented in Table \ref{tbl:spts}, along with the estimated reddening.  21 of our candidates observed with SpeX have spectral classifications available in the literature (see Table \ref{tbl:spts} for references).  The uncertainty of the \citet{allers13} spectral typing method is $\pm$1~subtype.  All but one of our spectral type determinations agree to within 1 subtype of an available literature spectral type, and all agree to within 1.5 subtypes.  Our estimate of reddening is likely to be very uncertain, as it depends on the reliability of synthetic photometry for our candidates and standards and assumes that our spectral standards represent intrinsic $J-K$ colors for young objects.  As noted in \citet{allers13}, however, a modest amount of reddening ($A_V$ $\lesssim$ 7~mags) does not affect spectral type or gravity determinations.  Figure \ref{fig:spectra} shows de-reddened spectra of our candidates having spectral types of M5 and later.

\section{The \wfilt-band Filter Confirmation Rate}

Of the 60 candidates for which we obtained spectra, 48 (80\%) are confirmed as having spectral types of M5 or later.  The 12 contaminant spectra are relatively featureless at SpeX prism resolution, and do not show spectral shapes that could result in a $Q$-value below -0.6.  Their spectra are consistent with being reddened background stars.  Most of our contaminants occur in regions of high extinction, where our \wfilt-band photometric calibration technique has high systematic uncertainty due to the relatively few number of background sources.  

Since the $Q$-values of young brown dwarfs and old field objects are similarly dependent on spectral type, the 48 confirmed late-M and L type objects are not necessarily members of the coincident star-forming regions.    
Using the method of \citet{allers13}, we also calculated gravity scores from our spectra.  Young objects in star-forming regions are expected to have gravity scores of {\sc    vl-g}.  Of the 48 confirmed late-type objects, 40 are classified as {\sc    vl-g}, 5 as {\sc    int-g}, and 1 as {\sc  fld-g}.  Two objects, with spectral types of M5, are unable to have their gravities classified using their low-resolution spectra and the method of \citet{allers13}.  

We also list the GAIA DR2 parallaxes and proper motions for candidate members in table \ref{tbl:spts}.  Of the 22 {\sc    vl-g}-classified spectra having GAIA parallaxes, 21 have parallaxes consistent (to within 2$\sigma$) with known members of their respective star-forming regions.    Of the three {\sc    int-g}-classified spectra having GAIA parallaxes, 2 have parallaxes consistent with their respective star-forming regions.  In total, we have 43 objects having spectral types of M5 and later with gravity classifications of {\sc    vl-g} or parallaxes consistent with membership, and 4 objects having late-type spectra but uncertain membership.  Overall, the confirmation rate from our SpeX survey is 80\% if we consider late-type objects, and 72\% if we consider only objects with gravity classifications or parallax consistent with membership in their natal star-forming regions.

The regions we surveyed with the \wfilt\ filter have long been known to contain brown dwarfs (see references in Table \ref{tbl:spts} for examples). When we created our target list for SpeX spectroscopic followup, we omitted 47 known brown dwarfs that met our \wfilt-band selection criteria.  Thus, if we had done a blind spectroscopic survey, we would have potentially confirmed those 47 known brown dwarfs (in addition to the 43 confirmed in our spectroscopic sample), which would imply an overall confirmation rate of 89\% (for objects later than M5) and 84\% for objects with gravity classifications or parallax consistent with membership. Our confirmation rate is comparable to the best performance ($\approx$80\%) achieved from high-quality surveys that use many more photometric bands \citep[e.g.,][]{2012ApJ...754...30P, 2017ApJ...842...65Z} and include proper-motion selection \citep[e.g.,][]{2018MNRAS.473.2020L}. For a survey with selection criteria determined by only 3 photometric bands, our confirmation rate is outstanding.

\section{Conclusions}
To search for brown dwarfs in star-forming regions, we have designed the \wfilt-band filter, a medium-bandpass filter centered at 1.45 $\mu$m.  Our \wfilt-band filter can be combined with broad-band $J$ and $H$ photometry to create a reddening-insensitive index, $Q$.  Since the \wfilt-band filter is centered on a deep atmospheric water absorption feature seen in the spectra of objects with spectral types of M5 and later, it is able to break the degeneracy in broad-band colors of young brown dwarfs and reddened background stars.  The combination of $J, H, $ and $W$ photometry can determine the spectral types of brown dwarfs to within 1.4 subtypes, independent of reddening.

We tested the efficacy of our filter by selecting candidate young brown dwarfs based only on their $Q$ values.  Using SpeX, we spectroscopically confirm 48/60 objects as having spectral types of M5 and later.  If we include objects that met our selection criteria but were already spectroscopically confirmed prior to our SpeX observations, our confirmation rate is 95/107.   The high confirmation rate of our survey demonstrates the ability of custom near-IR filters to identify young brown dwarfs.  Our spectroscopic confirmation includes seven objects with spectral types of L0 or later (3 of which are new discoveries), making them among the lowest mass confirmed members of star-forming regions.

Based on the success of our pilot survey with the $W$ filter, we subsequently collaborated on the installation of a similar filter on the Canada France Hawai'i Telescope's Wide-Field IR camera.  The first discoveries from this CFHT survey of Serpens have been recently reported \citep{jose20}, providing further confirmation of the effectiveness of the $W$-filter selection, even for star-forming regions near the Galactic Plane.

\acknowledgements
This publication makes use of data products from the Two Micron All Sky Survey, which is a joint project of the University of Massachusetts and the Infrared Processing and Analysis Center/California Institute of Technology, funded by the National Aeronautics and Space Administration and the National Science Foundation.  This research has benefited from the SpeX Prism Spectral Libraries, maintained by Adam Burgasser at \url{http://pono.ucsd.edu/~adam/browndwarfs/spexprism}.  We thank John Rayner and Shane Jacobson for assisting with design and installation of the $W$-band filter.  We are grateful to Geoff Mathews and Mark Pitts for conducting some of the SpeX observations.  Zhoujian Zhang and Michael Gully-Santiago provided useful conversations which improved this manuscript.  We are grateful to members of the \wfilt\ filter collaboration for sharing ideas on data reduction, analysis, and scientific results.    KNA thanks Greg Herczeg and the Kavli Institute for Astronomy and Astrophysics at Peking University for supporting a collaborative visit.  This work benefited from the 2019 Exoplanet Summer Program in the Other Worlds Laboratory (OWL) at the University of California, Santa Cruz, a program funded by the Heising-Simons Foundation.  This work was supported by NSF grant AST-0407441.  Support for program numbers 15238 and 15628 was provided through a grant from the STScI under NASA contract NAS5-26555.

\startlongtable
\begin{deluxetable}{llccl}
\tablecolumns{5}
\tabletypesize{\scriptsize}
\tablecaption{Observing Log: $W$ Band Imaging \label{tbl:wobs}}
\tablewidth{0pt}
\tablehead{
\colhead{Position} &
\colhead{Date} &
\colhead{$\sec z$} &
\colhead{$T_{\rm{int}}$}             &
\colhead{Seeing}             \\    
\colhead{} &
\colhead{(UT)} &
\colhead{} &
\colhead{(minutes)}             &
\colhead{(arcsec)}                
}
\startdata
 16 27 46.6  $-$24 25 38 & 2007 May 22 & 1.47 & 45.5 & 0.9\\ 
 16 27 46.6  $-$24 07 24 & 2007 May 22 & 1.39 & 45.5 & 1.2\\ 
 16 26 28.8  $-$24 25 24 & 2007 May 22 & 1.48 & 45.5 & 0.8\\ 
 16 26 28.8  $-$24 08 38 & 2007 May 22 & 1.67 & 45.5 & 1.2\\ 
 16 26 28.8  $-$24 43 34 & 2007 May 24 & 1.46 & 73.5 & 1.0\\ 
 16 27 46.6  $-$24 42 30 & 2007 May 24 & 1.53 & 91.0 & 0.9\\ 
 16 23 03.0  $-$23 37 56 & 2007 May 24 & 2.00 & 31.5 & 1.2\\ 
 16 23 03.0  $-$23 39 14 & 2007 May 26 & 1.50 & 42.0 & 1.3\\ 
 16 23 03.3  $-$24 12 20 & 2007 May 26 & 1.53 & 56.0 & 1.5\\ 
 16 23 03.2  $-$23 55 10 & 2007 May 26 & 1.55 & 80.5 & 1.5\\ 
 16 21 46.0  $-$23 56 56 & 2007 May 27 & 1.59 & 45.5 & 1.2\\ 
 16 22 07.0  $-$23 31 44 & 2007 May 27 & 1.40 & 45.5 & 1.4\\ 
 16 22 07.0  $-$23 14 41 & 2007 May 27 & 1.46 & 84.0 & 1.1\\ 
 16 29 04.3  $-$24 42 34 & 2007 May 27 & 1.89 & 45.5 & 1.2\\ 
 16 29 04.3  $-$24 26 34 & 2007 May 29 & 1.75 & 45.5 & 1.2\\ 
 16 31 54.0  $-$24 47 39 & 2007 May 29 & 1.50 & 45.5 & 1.2\\ 
 16 33 12.0  $-$24 48 20 & 2007 May 29 & 1.41 & 45.5 & 1.3\\ 
 16 31 54.0  $-$24 29 57 & 2007 May 29 & 1.45 & 45.5 & 1.2\\ 
 16 33 12.0  $-$24 28 41 & 2007 May 29 & 1.73 & 63.0 & 1.1\\ 
 16 33 12.2  $-$24 12 39 & 2007 Jun 3 & 1.70 & 42.0 & 1.2\\ 
 16 31 54.0  $-$24 11 57 & 2007 Jun 3 & 1.48 & 45.5 & 1.1\\ 
 16 34 30.0  $-$24 48 09 & 2007 Jun 3 & 1.41 & 45.5 & 1.3\\ 
 16 34 30.8  $-$24 29 03 & 2007 Jun 3 & 1.44 & 28.0 & 1.1\\ 
 16 34 30.0  $-$24 11 11 & 2007 Jun 3 & 1.64 & 45.5 & 1.1\\ 
 16 21 46.0  $-$24 14 16 & 2007 Jun 23 & 1.47 & 45.5 & 1.2\\ 
 16 20 51.6  $-$23 15 00 & 2007 Jun 23 & 1.37 & 42.0 & 1.2\\ 
 16 20 52.0  $-$23 32 09 & 2007 Jun 23 & 1.46 & 45.5 & 1.2\\ 
 16 21 46.0  $-$23 39 02 & 2007 Jun 23 & 1.74 & 45.5 & 1.3\\ 
 16 26 28.8  $-$24 47 07 & 2008 Jun 13 & 1.48 & 63.0 & 1.5\\ 
  16 29 04.3  $-$24 11 25 & 2008 Jun 13 & 1.41 & 59.5 & 1.7\\ 
 16 25 19.0  $-$24 48 23 & 2008 Jun 13 & 1.65 & 45.5 & 1.7\\ 
 16 25 19.0  $-$24 39 15 & 2008 Jun 14 & 1.62 & 42.0 & 1.2\\ 
 16 25 19.0  $-$24 30 18 & 2008 Jun 14 & 1.44 & 45.5 & 1.3\\ 
  16 25 19.0  $-$24 21 25 & 2008 Jun 14 & 1.40 & 45.5 & 1.2\\ 
 16 25 19.0  $-$24 12 39 & 2008 Jun 14 & 1.51 & 45.5 & 1.2\\ 
 16 25 19.0  $-$24 04 16 & 2008 Jun 14 & 1.82 & 45.5 & 1.8\\ 
 16 24 06.0  $-$24 48 16 & 2008 Jun 18 & 1.66 & 35.0 & 1.2\\ 
 16 24 06.0  $-$24 30 52 & 2008 Jun 18 & 1.46 & 45.5 & 1.4\\ 
 16 24 06.0  $-$24 12 52 & 2008 Jun 18 & 1.40 & 45.5 & 1.1\\ 
 16 24 06.0  $-$23 55 09 & 2008 Jun 18 & 1.49 & 45.5 & 1.0\\ 
 16 24 06.0  $-$24 04 15 & 2008 Jun 18 & 1.67 & 17.5 & 1.3\\ 
 16 24 06.0  $-$24 39 28 & 2008 Jun 19 & 1.65 & 38.5 & 1.2\\ 
 16 24 06.0  $-$24 21 56 & 2008 Jun 19 & 1.45 & 45.5 & 1.2\\ 
 03 41 55.0  +31 33 34 & 2008 Aug 15 & 1.20 & 45.5 & 1.3\\ 
 03 43 10.0  +31 33 37 & 2008 Aug 15 & 1.09 & 45.5 & 1.0\\ 
 03 44 25.0  +31 33 02 & 2008 Aug 17 & 1.44 & 45.5 & 1.1\\ 
 03 45 40.0  +31 33 34 & 2008 Aug 17 & 1.19 & 45.5 & 1.1\\ 
 03 41 55.0  +31 49 32 & 2008 Aug 17 & 1.07 & 45.5 & 1.0\\ 
 03 28 50.0  +31 07 54 & 2008 Oct 19 & 1.66 & 45.5 & 0.9\\ 
 03 30 00.0  +31 07 46 & 2008 Oct 19 & 1.34 & 45.5 & 1.1\\ 
 03 28 50.0  +31 17 29 & 2008 Oct 19 & 1.15 & 45.5 & 0.9\\ 
 03 30 00.0  +31 17 43 & 2008 Oct 19 & 1.06 & 45.5 & 1.1\\ 
 03 31 10.0  +31 08 05 & 2008 Oct 19 & 1.02 & 45.5 & 1.1\\ 
 03 31 10.0  +31 17 39 & 2008 Oct 19 & 1.04 & 45.5 & 1.0\\ 
 03 28 50.0  +30 58 42 & 2008 Oct 19 & 1.12 & 45.5 & 1.0\\ 
 03 30 00.0  +30 58 44 & 2008 Oct 19 & 1.27 & 45.5 & 1.0\\ 
 03 31 10.2  +30 58 46 & 2008 Oct 19 & 1.51 & 42.0 & 1.1\\ 
03 43 09.0  +31 55 34 & 2008 Nov 6 & 1.51 & 45.5 & 0.7\\ 
03 44 26.0  +31 55 59 & 2008 Nov 6 & 1.25 & 45.5 & 0.7\\ 
03 45 44.0  +31 55 47 & 2008 Nov 6 & 1.10 & 45.5 & 0.7\\ 
03 43 09.0  +32 06 00 & 2008 Nov 6 & 1.74 & 56.0 & 0.7\\ 
03 44 26.0  +32 04 54 & 2008 Nov 6 & 1.40 & 45.5 & 0.7\\ 
03 45 44.0  +32 04 55 & 2008 Nov 6 & 1.18 & 45.5 & 0.7\\ 
03 43 09.0  +32 14 14 & 2008 Nov 6 & 1.06 & 45.5 & 0.8\\ 
03 44 26.0  +32 14 27 & 2008 Nov 6 & 1.03 & 45.5 & 0.8\\ 
03 45 44.0  +32 14 40 & 2008 Nov 6 & 1.04 & 45.5 & 0.8\\ 
03 43 09.0  +31 44 10 & 2008 Nov 15 & 1.50 & 45.5 & 0.7\\ 
03 44 26.0  +31 45 29 & 2008 Nov 15 & 1.21 & 45.5 & 0.7\\ 
03 45 43.6  +31 45 50 & 2008 Nov 15 & 1.09 & 42.0 & 0.7\\ 
03 43 27.0  +32 24 20 & 2008 Nov 15 & 1.03 & 45.5 & 0.8\\ 
03 44 37.0  +32 23 59 & 2008 Nov 15 & 1.03 & 45.5 & 0.8\\ 
03 45 47.0  +32 24 09 & 2008 Nov 15 & 1.08 & 45.5 & 0.9\\ 
03 28 50.0  +30 52 04 & 2008 Nov 15 & 1.25 & 45.5 & 0.7\\ 
03 30 00.0  +30 51 56 & 2008 Nov 15 & 1.49 & 45.5 & 0.7\\ 
03 31 10.0  +30 48 40 & 2008 Nov 16 & 1.56 & 45.5 & 1.1\\ 
03 28 50.0  +30 39 39 & 2008 Nov 16 & 1.24 & 45.5 & 1.1\\ 
03 30 00.0  +30 39 40 & 2008 Nov 16 & 1.10 & 45.5 & 1.1\\ 
03 31 10.0  +30 39 33 & 2008 Nov 16 & 1.03 & 45.5 & 1.2\\ 
03 28 50.0  +30 30 24 & 2008 Nov 16 & 1.02 & 45.5 & 1.1\\ 
03 30 00.0  +30 30 32 & 2008 Nov 16 & 1.07 & 45.5 & 1.2\\ 
03 31 10.0  +30 30 27 & 2008 Nov 16 & 1.17 & 45.5 & 1.1\\ 
\enddata
\end{deluxetable}

\newpage

\startlongtable
\begin{deluxetable}{lcccccccc}
\tablecaption{Photometry of Observed Brown Dwarf Candidates \label{tbl:candphot}}
\tabletypesize{\scriptsize}
\tablewidth{0pt}
\tablehead{
  \colhead{Object} &
  \colhead{$W$} &
  \colhead{$J_\mathrm{MKO}$} &
  \colhead{$H_\mathrm{MKO}$} &
  \colhead{$J_\mathrm{2MASS}$} &
  \colhead{$H_\mathrm{2MASS}$} &
  \colhead{$Q_\mathrm{MKO}$} &
  \colhead{$Q_\mathrm{2MASS}$} \\
  \colhead{} &
  \colhead{(mag)} &
  \colhead{(mag)} &
  \colhead{(mag)} &
  \colhead{(mag)} &
  \colhead{} &
  \colhead{} 
} 
\startdata
\multicolumn{8}{c}{NGC~1333} \\           
\tableline            
\\
UHW~J052.25592+30.82075 & 13.51$\pm$0.02 & 13.49$\pm$0.01 & 12.99$\pm$0.01 & 13.53$\pm$0.03 & 12.89$\pm$0.03 & -0.86$\pm$0.05 & -0.92$\pm$0.07 \\
UHW~J052.26618+30.94156 & 15.82$\pm$0.04 & 15.94$\pm$0.01 & 15.15$\pm$0.01 & 15.64$\pm$0.06 & 14.64$\pm$0.07 & -0.96$\pm$0.10 & -1.98$\pm$0.16 \\
UHW~J052.27886+31.49911 & 16.28$\pm$0.05 & 16.33$\pm$0.01 & 15.53$\pm$0.01 & 16.68$\pm$0.05 & 15.52$\pm$0.05 & -1.16$\pm$0.13 & -0.76$\pm$0.15 \\
UHW~J052.30643+31.39637 & 17.09$\pm$0.05 & 17.04$\pm$0.02 & 16.18$\pm$0.01 &      \nodata          & 16.10$\pm$0.07 & -1.51$\pm$0.14 &      \nodata          \\
UHW~J052.35191+31.47077 & 13.84$\pm$0.04 & 13.88$\pm$0.01 & 13.18$\pm$0.01 & 14.06$\pm$0.02 & 13.19$\pm$0.02 & -1.03$\pm$0.11 & -0.78$\pm$0.11 \\
UHW~J052.40147+31.29717 & 17.92$\pm$0.05 & 17.77$\pm$0.01 & 17.15$\pm$0.01 & 17.92$\pm$0.15 & 17.12$\pm$0.22 & -1.39$\pm$0.12 & -1.22$\pm$0.39 \\
UHW~J052.43392+31.32982 & 17.40$\pm$0.03 & 17.58$\pm$0.01 & 16.52$\pm$0.01 & 17.45$\pm$0.09 &      \nodata          & -1.24$\pm$0.08 &      \nodata          \\
UHW~J052.75599+30.63090 & 16.43$\pm$0.03 & 16.30$\pm$0.01 & 15.60$\pm$0.01 & 16.57$\pm$0.16 & 15.85$\pm$0.18 & -1.47$\pm$0.08 & -0.76$\pm$0.34 \\
\tableline
\multicolumn{8}{c}{IC~348} \\
\tableline
\\
UHW~J055.48646+31.83838 & 15.08$\pm$0.02 & 14.97$\pm$0.01 & 14.42$\pm$0.01 & 15.06$\pm$0.04 & 14.36$\pm$0.05 & -1.16$\pm$0.05 & -1.10$\pm$0.10 \\
UHW~J055.54657+31.94999 & 15.66$\pm$0.04 & 15.90$\pm$0.01 & 14.63$\pm$0.01 & 16.10$\pm$0.09 & 14.65$\pm$0.06 & -1.42$\pm$0.10 & -1.10$\pm$0.16 \\
UHW~J055.55875+31.95973 & 14.53$\pm$0.02 & 14.85$\pm$0.01 & 13.55$\pm$0.01 & 14.97$\pm$0.04 & 13.57$\pm$0.04 & -1.24$\pm$0.06 & -1.01$\pm$0.09 \\
UHW~J055.67611+31.72142 & 16.37$\pm$0.03 & 16.65$\pm$0.03 & 15.64$\pm$0.01 & 16.77$\pm$0.13 & 15.49$\pm$0.10 & -0.90$\pm$0.08 & -0.94$\pm$0.21 \\
UHW~J055.71237+31.94127 & 16.37$\pm$0.04 & 16.44$\pm$0.01 & 15.65$\pm$0.01 & 16.59$\pm$0.11 & 15.54$\pm$0.12 & -1.09$\pm$0.11 & -1.03$\pm$0.25 \\
UHW~J055.73277+32.07455 & 14.73$\pm$0.03 & 14.61$\pm$0.01 & 13.94$\pm$0.01 & 14.76$\pm$0.02 & 13.99$\pm$0.02 & -1.38$\pm$0.07 & -1.08$\pm$0.08 \\
UHW~J055.75603+32.03731 & 14.34$\pm$0.02 & 14.40$\pm$0.01 & 13.67$\pm$0.01 & 14.49$\pm$0.02 & 13.68$\pm$0.02 & -1.02$\pm$0.06 & -0.86$\pm$0.07 \\
UHW~J055.75810+31.74320 & 14.77$\pm$0.03 & 14.82$\pm$0.01 & 13.90$\pm$0.01 & 14.91$\pm$0.04 & 13.93$\pm$0.04 & -1.34$\pm$0.08 & -1.13$\pm$0.11 \\
UHW~J055.76964+32.16007 & 16.90$\pm$0.03 & 16.95$\pm$0.01 & 16.12$\pm$0.01 & 17.06$\pm$0.06 & 16.03$\pm$0.07 & -1.20$\pm$0.08 & -1.15$\pm$0.15 \\
UHW~J055.98976+32.24691 & 14.83$\pm$0.02 & 14.89$\pm$0.01 & 14.23$\pm$0.01 & 14.96$\pm$0.02 & 14.23$\pm$0.03 & -0.91$\pm$0.05 & -0.79$\pm$0.06 \\
UHW~J056.01221+31.87436 & 14.23$\pm$0.02 & 14.46$\pm$0.01 & 13.51$\pm$0.03 & 14.71$\pm$0.03 & 13.57$\pm$0.03 & -0.94$\pm$0.05 & -0.53$\pm$0.06 \\
UHW~J056.04265+31.54385 & 16.15$\pm$0.03 & 16.15$\pm$0.01 & 15.51$\pm$0.01 & 16.29$\pm$0.11 & 15.33$\pm$0.09 & -1.03$\pm$0.07 & -1.11$\pm$0.19 \\
UHW~J056.07487+31.58792 & 16.83$\pm$0.03 & 17.04$\pm$0.01 & 15.87$\pm$0.01 & 16.69$\pm$0.15 & 15.91$\pm$0.16 & -1.33$\pm$0.07 & -1.53$\pm$0.30 \\
UHW~J056.14403+31.71225 & 14.83$\pm$0.02 & 15.07$\pm$0.03 & 14.08$\pm$0.03 & 15.23$\pm$0.05 & 14.08$\pm$0.03 & -0.96$\pm$0.08 & -0.74$\pm$0.09 \\
UHW~J056.17321+32.17758 & 15.17$\pm$0.03 & 15.25$\pm$0.01 & 14.62$\pm$0.01 & 15.39$\pm$0.03 & 14.68$\pm$0.03 & -0.81$\pm$0.07 & -0.52$\pm$0.09 \\
UHW~J056.34957+32.19849 & 15.09$\pm$0.02 & 15.26$\pm$0.04 & 14.52$\pm$0.03 & 15.33$\pm$0.02 & 14.48$\pm$0.03 & -0.74$\pm$0.06 & -0.68$\pm$0.05 \\
UHW~J056.37556+31.62740 & 16.25$\pm$0.02 & 16.27$\pm$0.01 & 15.36$\pm$0.01 & 16.43$\pm$0.13 & 15.26$\pm$0.08 & -1.41$\pm$0.05 & -1.33$\pm$0.18 \\
UHW~J056.38941+31.76537 & 14.96$\pm$0.03 & 15.12$\pm$0.01 & 14.04$\pm$0.01 & 15.26$\pm$0.05 & 14.00$\pm$0.04 & -1.32$\pm$0.07 & -1.17$\pm$0.10 \\
\tableline
\multicolumn{8}{c}{Ophiuchus} \\
\tableline
UHW~J245.25929$-$23.97766 & 14.40$\pm$0.02 & 14.63$\pm$0.01 & 13.80$\pm$0.01 & 14.75$\pm$0.04 & 13.79$\pm$0.03 & -0.76$\pm$0.05 & -0.59$\pm$0.05 \\
UHW~J245.32623$-$23.90277 & 17.22$\pm$0.02 & 17.60$\pm$0.04 & 16.42$\pm$0.02 &      \nodata          &      \nodata          & -0.92$\pm$0.08 &      \nodata          \\
UHW~J245.39963$-$23.91766 & 13.82$\pm$0.02 & 13.79$\pm$0.01 & 13.16$\pm$0.01 & 13.95$\pm$0.03 & 13.19$\pm$0.02 & -1.09$\pm$0.05 & -0.82$\pm$0.05 \\
UHW~J245.48320$-$23.35050 & 11.48$\pm$0.02 &      \nodata          &      \nodata          & 11.50$\pm$0.03 & 10.50$\pm$0.02 &      \nodata          & -1.47$\pm$0.05 \\
UHW~J245.58499$-$23.13122 & 16.59$\pm$0.02 &      \nodata          &      \nodata          & 16.95$\pm$0.18 & 15.42$\pm$0.12 &      \nodata          & -1.42$\pm$0.27 \\
UHW~J245.66849$-$23.56412 & 12.32$\pm$0.02 &      \nodata          &      \nodata          & 12.72$\pm$0.02 & 11.48$\pm$0.02 &      \nodata          & -0.88$\pm$0.05 \\
UHW~J245.68198$-$24.03424 & 17.04$\pm$0.03 & 17.32$\pm$0.01 & 16.30$\pm$0.01 &      \nodata          &      \nodata          & -0.90$\pm$0.07 &      \nodata          \\
UHW~J245.71139$-$24.22963 & 15.39$\pm$0.02 & 15.47$\pm$0.01 & 14.92$\pm$0.01 & 15.46$\pm$0.05 & 14.83$\pm$0.07 & -0.67$\pm$0.05 & -0.78$\pm$0.12 \\
UHW~J245.81343$-$23.78480 & 14.98$\pm$0.02 & 15.10$\pm$0.01 & 14.21$\pm$0.01 & 15.19$\pm$0.04 & 14.20$\pm$0.05 & -1.13$\pm$0.06 & -0.97$\pm$0.11 \\
UHW~J245.87762$-$24.22221 & 11.76$\pm$0.02 &      \nodata          &      \nodata          & 11.93$\pm$0.02 & 11.01$\pm$0.02 &      \nodata          & -0.97$\pm$0.05 \\
UHW~J245.95168$-$24.22649 & 13.04$\pm$0.02 &      \nodata          &      \nodata          & 13.53$\pm$0.03 & 12.17$\pm$0.03 &      \nodata          & -0.84$\pm$0.05 \\
UHW~J246.06631$-$24.18753 & 13.48$\pm$0.05 & 13.83$\pm$0.01 & 12.76$\pm$0.01 & 13.97$\pm$0.03 & 12.79$\pm$0.02 & -0.81$\pm$0.13 & -0.57$\pm$0.13 \\
UHW~J246.51367$-$24.50719 & 17.50$\pm$0.06 & 17.85$\pm$0.02 & 16.56$\pm$0.02 & 17.68$\pm$0.17 & 16.18$\pm$0.10 & -1.16$\pm$0.16 & -1.83$\pm$0.28 \\
UHW~J246.57742$-$24.49763 & 16.41$\pm$0.03 & 17.23$\pm$0.02 & 15.11$\pm$0.02 & 17.36$\pm$0.13 & 15.01$\pm$0.04 & -1.26$\pm$0.08 & -1.18$\pm$0.16 \\
UHW~J246.62661$-$24.20326 & 15.05$\pm$0.02 & 16.18$\pm$0.02 & 13.83$\pm$0.02 & 16.32$\pm$0.05 & 13.80$\pm$0.02 & -0.85$\pm$0.05 & -0.65$\pm$0.06 \\
UHW~J246.65753$-$24.65089 & 14.41$\pm$0.02 & 14.85$\pm$0.02 & 13.68$\pm$0.01 & 14.93$\pm$0.03 & 13.59$\pm$0.02 & -0.75$\pm$0.05 & -0.73$\pm$0.05 \\
UHW~J246.68094$-$24.17893 & 13.31$\pm$0.02 &      \nodata          &      \nodata          & 13.60$\pm$0.02 & 12.68$\pm$0.02 &      \nodata          & -0.66$\pm$0.05 \\
UHW~J246.71774$-$24.02967 & 15.72$\pm$0.02 & 16.08$\pm$0.01 & 14.90$\pm$0.01 & 16.08$\pm$0.04 & 14.83$\pm$0.04 & -0.97$\pm$0.05 & -1.00$\pm$0.09 \\
UHW~J246.73470$-$24.71048 & 17.45$\pm$0.06 & 17.56$\pm$0.02 & 16.62$\pm$0.02 & 17.77$\pm$0.19 & 16.59$\pm$0.14 & -1.24$\pm$0.15 & -0.99$\pm$0.33 \\
UHW~J246.82722$-$24.12273 & 14.30$\pm$0.02 & 14.73$\pm$0.01 & 13.58$\pm$0.01 & 14.83$\pm$0.02 & 13.55$\pm$0.02 & -0.72$\pm$0.05 & -0.60$\pm$0.05 \\
UHW~J246.90062$-$24.08269 & 13.74$\pm$0.02 & 14.11$\pm$0.02 & 12.91$\pm$0.03 & 14.29$\pm$0.02 & 12.95$\pm$0.02 & -0.97$\pm$0.06 & -0.65$\pm$0.05 \\
UHW~J246.92017$-$24.48353 & 14.53$\pm$0.02 & 14.64$\pm$0.02 & 13.86$\pm$0.02 & 14.71$\pm$0.02 & 13.78$\pm$0.02 & -0.98$\pm$0.05 & -0.96$\pm$0.05 \\
UHW~J246.94784$-$24.82865 & 16.38$\pm$0.04 &      \nodata          &      \nodata          & 16.56$\pm$0.05 & 15.51$\pm$0.05 &      \nodata          & -1.16$\pm$0.15 \\
UHW~J246.99312$-$23.98747 & 13.95$\pm$0.02 & 14.11$\pm$0.02 & 13.28$\pm$0.02 & 14.22$\pm$0.02 & 13.20$\pm$0.02 & -0.92$\pm$0.05 & -0.87$\pm$0.05 \\
UHW~J247.01994$-$24.84319 & 13.86$\pm$0.02 & 13.91$\pm$0.02 & 13.09$\pm$0.01 & 13.94$\pm$0.02 & 12.98$\pm$0.02 & -1.18$\pm$0.05 & -1.24$\pm$0.05 \\
UHW~J247.04356$-$24.40568 & 17.32$\pm$0.04 & 17.78$\pm$0.02 & 16.28$\pm$0.01 & \nodata & 16.19$\pm$0.09 & -1.22$\pm$0.11 & \nodata \\
UHW~J247.19598$-$24.47051 & 13.35$\pm$0.02 &      \nodata          &      \nodata          & 13.24$\pm$0.03 & 12.53$\pm$0.02 &      \nodata          & -1.35$\pm$0.05 \\
UHW~J247.20293$-$24.44216 & 14.35$\pm$0.02 &      \nodata          &      \nodata          & 14.45$\pm$0.03 & 13.67$\pm$0.03 &      \nodata          & -0.95$\pm$0.07 \\
UHW~J247.20293$-$24.55216 & 16.56$\pm$0.02 & 17.19$\pm$0.01 & 16.08$\pm$0.01 &      \nodata          &      \nodata          & -0.13$\pm$0.06 &      \nodata          \\
UHW~J247.38539$-$24.04299 & 16.50$\pm$0.03 &      \nodata          &      \nodata          & 16.82$\pm$0.15 & 15.59$\pm$0.10 &      \nodata          & -1.07$\pm$0.24 \\
UHW~J247.94088$-$24.62855 & 12.73$\pm$0.07 &    \nodata          &      \nodata          & 13.09$\pm$0.02 & 11.85$\pm$0.02 &     \nodata          &  -0.98$\pm$0.17  \\ 
UHW~J247.95427$-$24.78060 & 17.55$\pm$0.08 & 17.65$\pm$0.01 & 16.61$\pm$0.03 &      \nodata          &      \nodata          & -1.42$\pm$0.21 &      \nodata          \\
UHW~J248.69073$-$24.82275 & 16.05$\pm$0.02 & 16.25$\pm$0.01 & 15.50$\pm$0.02 & 16.09$\pm$0.09 & 15.27$\pm$0.09 & -0.70$\pm$0.05 & -1.15$\pm$0.18 \\
\enddata
\end{deluxetable}

\newpage

\startlongtable
\begin{deluxetable}{lllllllll}
\tablecaption{IRTF/SpeX Observing Log \label{tbl:spex_observing}}
\tabletypesize{\scriptsize}
\tablewidth{0pt}
\tablehead{
 \colhead{Object} &
 \colhead{UT Date} &
 \colhead{Grat} &
 \colhead{Slit (\arcsec)} &
 \colhead{$R$} &
 \colhead{$\sec z$} &
 \colhead{$N_{\rm{exp}} \times t$ (s)} &
 \colhead{$T_{\rm{int}}$ (s)} &
 \colhead{$<S/N> (Y,J,H,K)$} 
 }
\startdata
UHW~J052.26618+30.94156 & 2009 Sep 30   &  LowRes15  &   0.5$\times$15 &    120 &   1.25 &   10$\times$120.0 &   1200.0 &   30,  45,  45,  40 \\
UHW~J052.27886+31.49911 & 2009 Sep 29   &  LowRes15  &   0.5$\times$15 &    120 &   1.29 &   10$\times$180.0 &   1800.0 &   47, 102,  95,  85 \\
UHW~J052.30643+31.39637 & 2009 Sep 30   &  LowRes15  &   0.5$\times$15 &    120 &   1.10 &   12$\times$180.0 &   2160.0 &    2,   7,  19,  30 \\
UHW~J052.35191+31.47077 & 2009 Oct 1   &  LowRes15  &   0.5$\times$15 &    120 &   1.38 &   10$\times$60.0  &    600.0 &  135, 255, 244, 181 \\
UHW~J052.40147+31.29717 & 2009 Sep 28   &  LowRes15  &   0.8$\times$15 &     75 &   1.09 &   26$\times$180.0 &   4680.0 &   30,  62,  46,  43 \\
UHW~J052.43392+31.32982 & 2009 Sep 29   &  LowRes15  &   0.5$\times$15 &    120 &   1.11 &   12$\times$180.0 &   2160.0 &   21,  37,  42,  63 \\
UHW~J052.75599+30.63090 & 2009 Sep 28   &  LowRes15  &   0.5$\times$15 &    120 &   1.02 &   18$\times$120.0 &   2160.0 &   37,  76,  60,  48 \\
UHW~J055.48646+31.83838 & 2009 Oct 1   &  LowRes15  &   0.5$\times$15 &    120 &   1.27 &   10$\times$120.0 &   1200.0 &  117, 168, 124,  79 \\
UHW~J055.54657+31.94999 & 2009 Sep 28   &  LowRes15  &   0.5$\times$15 &    120 &   1.10 &   11$\times$60.0  &    660.0 &   17,  45,  59,  44 \\
UHW~J055.55875+31.95973 & 2009 Sep 27   &  LowRes15  &   0.5$\times$15 &    120 &   1.13 &   10$\times$60.0  &    600.0 &   35, 108, 168, 118 \\
UHW~J055.67611+31.72142 & 2009 Oct 2   &  LowRes15  &   0.5$\times$15 &    120 &   1.15 &   16$\times$180.0 &   2880.0 &   14,  37,  36,  37 \\
UHW~J055.71237+31.94127 & 2009 Oct 2   &  LowRes15  &   0.5$\times$15 &    120 &   1.03 &   16$\times$180.0 &   2880.0 &   31,  62,  47,  41 \\
UHW~J055.73277+32.07455 & 2009 Sep 27   &  LowRes15  &   0.5$\times$15 &    120 &   1.04 &   10$\times$60.0  &    600.0 &  126, 234, 215, 148 \\
UHW~J055.75603+32.03731 & 2009 Oct 2   &  LowRes15  &   0.5$\times$15 &    120 &   1.16 &   11$\times$60.0  &    660.0 &   98, 193, 165, 134 \\
UHW~J055.75810+31.74320 & 2009 Sep 27   &  LowRes15  &   0.5$\times$15 &    120 &   1.39 &   10$\times$120.0 &   1200.0 &  103, 228, 210, 194 \\
UHW~J055.76964+32.16007 & 2009 Sep 27   &  LowRes15  &   0.5$\times$15 &    120 &   1.06 &   13$\times$180.0 &   2340.0 &   28,  66,  66,  47 \\
UHW~J055.98976+32.24691 & 2009 Sep 28   &  LowRes15  &   0.5$\times$15 &    120 &   1.40 &   10$\times$60.0  &    600.0 &   85, 154, 123,  90 \\
UHW~J056.01221+31.87436 & 2009 Oct 2   &  LowRes15  &   0.5$\times$15 &    120 &   1.32 &   10$\times$90.0  &    900.0 &   63, 144, 148, 137 \\
UHW~J056.04265+31.54385 & 2009 Sep 27   &  LowRes15  &   0.5$\times$15 &    120 &   1.02 &   10$\times$180.0 &   1800.0 &   63, 118,  95,  59 \\
UHW~J056.07487+31.58792 & 2009 Sep 28   &  LowRes15  &   0.5$\times$15 &    120 &   1.20 &   12$\times$180.0 &   2160.0 &   16,  35,  43,  33 \\
UHW~J056.14403+31.71225 & 2009 Oct 1   &  LowRes15  &   0.5$\times$15 &    120 &   1.15 &   16$\times$120.0 &   1920.0 &   43, 114, 118, 115 \\
UHW~J056.17321+32.17758 & 2009 Oct 2   &  LowRes15  &   0.5$\times$15 &    120 &   1.08 &   12$\times$120.0 &   1440.0 &   96, 162, 122,  97 \\
UHW~J056.34957+32.19849 & 2009 Sep 28   &  LowRes15  &   0.5$\times$15 &    120 &   1.08 &   10$\times$60.0  &    600.0 &   60, 120, 103,  79 \\
UHW~J056.37556+31.62740 & 2009 Sep 27   &  LowRes15  &   0.5$\times$15 &    120 &   1.08 &    9$\times$180.0 &   1620.0 &   41,  76,  75,  50 \\
UHW~J056.38941+31.76537 & 2009 Sep 27   &  LowRes15  &   0.5$\times$15 &    120 &   1.19 &   10$\times$120.0 &   1200.0 &   67, 167, 195, 178 \\
UHW~J245.25929$-$23.97766 & 2012 Jul 5   &  LowRes15  &   0.5$\times$15 &    120 &   1.75 &    6$\times$60.0  &    360.0 &   21,  41,  36,  30 \\
UHW~J245.32623$-$23.90277 & 2009 May 3   &  LowRes15  &   0.8$\times$15 &     75 &   1.46 &   10$\times$180.0 &   1800.0 &   22,  26,  19,  12 \\
UHW~J245.39963$-$23.91766 & 2008 Jul 2   &  LowRes15  &   0.5$\times$15 &    120 &   1.43 &   11$\times$30.0  &    330.0 &  147, 254, 231, 162 \\
UHW~J245.48320$-$23.35050 & 2012 Jul 6   &  LowRes15  &   0.3$\times$15 &    200 &   1.54 &    9$\times$15.0  &    135.0 &  147, 257, 279, 191 \\
UHW~J245.58499$-$23.13122 & 2009 May 3   &  LowRes15  &   0.5$\times$15 &    120 &   1.73 &    8$\times$180.0 &   1440.0 &   14,  33,  31,  34 \\
UHW~J245.66849$-$23.56412 & 2012 Jul 5   &  LowRes15  &   0.5$\times$15 &    120 &   1.64 &    7$\times$10.0  &     70.0 &   24,  63,  82,  63 \\
UHW~J245.68198$-$24.03424 & 2009 May 2   &  LowRes15  &   0.8$\times$15 &     75 &   1.61 &    8$\times$300.0 &   2400.0 &   20,  39,  33,  37 \\
UHW~J245.71139$-$24.22963 & 2009 May 6   &  LowRes15  &   0.8$\times$15 &     75 &   1.41 &    8$\times$180.0 &   1440.0 &   56,  67,  43,  37 \\
UHW~J245.81343$-$23.78480 & 2008 Jul 2   &  LowRes15  &   0.5$\times$15 &    120 &   1.48 &   11$\times$60.0  &    660.0 &   68, 156, 159, 142 \\
UHW~J245.87762$-$24.22221 & 2012 Jul 6   &  LowRes15  &   0.3$\times$15 &    200 &   1.59 &    7$\times$15.0  &    105.0 &   93, 183, 181, 136 \\
UHW~J245.95168$-$24.22649 & 2012 Jul 5   &  LowRes15  &   0.5$\times$15 &    120 &   1.59 &    6$\times$30.0  &    180.0 &   31,  85,  96,  63 \\
UHW~J246.06631$-$24.18753 & 2012 Jul 6   &  LowRes15  &   0.3$\times$15 &    200 &   1.63 &    6$\times$30.0  &    180.0 &   24,  63,  66,  45 \\
UHW~J246.51367$-$24.50719 & 2008 Jul 2   &  LowRes15  &   0.5$\times$15 &    120 &   1.41 &   15$\times$180.0 &   2700.0 &    4,  18,  28,  38 \\
UHW~J246.57742$-$24.49763 & 2008 Jul 2   &  LowRes15  &   0.5$\times$15 &    120 &   1.48 &   10$\times$180.0 &   1800.0 &    7,  41, 131, 225 \\
UHW~J246.62661$-$24.20326 & 2008 Jul 6   &  LowRes15  &   0.5$\times$15 &    120 &   1.76 &    4$\times$180.0 &    720.0 &    8,  55, 182, 306 \\
UHW~J246.65753$-$24.65089 & 2007 Jun 6   &   ShortXD  &   0.8$\times$15 &    750 &   1.50 &   12$\times$180.0 &   2160.0 &   17,  67, 109, 155 \\
UHW~J246.68094$-$24.17893 & 2008 Jul 3   &  LowRes15  &   0.5$\times$15 &    120 &   1.85 &   10$\times$60.0  &    600.0 &  174, 327, 292, 202 \\
UHW~J246.71774$-$24.02967 & 2008 Jun 27   &  LowRes15  &   0.5$\times$15 &    120 &   1.45 &    7$\times$180.0 &   1260.0 &   19,  54,  64,  71 \\
UHW~J246.73470$-$24.71048 & 2009 May 5   &  LowRes15  &   0.8$\times$15 &     75 &   1.43 &   12$\times$180.0 &   2160.0 &   16,  41,  36,  45 \\
UHW~J246.82722$-$24.12273 & 2008 Jul 3   &  LowRes15  &   0.5$\times$15 &    120 &   1.45 &   10$\times$60.0  &    600.0 &   82, 204, 239, 203 \\
UHW~J246.90062$-$24.08269 & 2007 Jun 7   &   ShortXD  &   0.8$\times$15 &    750 &   1.39 &   12$\times$90.0  &   1080.0 &    0,  48,  87, 132 \\
UHW~J246.92017$-$24.48353 & 2008 Jun 25   &   ShortXD  &   0.5$\times$15 &   1200 &   1.47 &   10$\times$300.0 &   3000.0 &   21,  61,  83, 106 \\
UHW~J246.94784$-$24.82865 & 2008 Jul 3   &  LowRes15  &   0.5$\times$15 &    120 &   1.61 &   13$\times$180.0 &   2340.0 &   34,  58,  51,  46 \\
UHW~J246.99312$-$23.98747 & 2008 Jul 2   &  LowRes15  &   0.5$\times$15 &    120 &   1.59 &   10$\times$30.0  &    300.0 &   93, 197, 204, 185 \\
UHW~J247.01994$-$24.84319 & 2008 Jun 27   &  LowRes15  &   0.5$\times$15 &    120 &   1.42 &   10$\times$15.0  &    150.0 &   40, 104, 105,  90 \\
UHW~J247.04356$-$24.40568 & 2007 Jun 5   &  LowRes15  &   0.5$\times$15 &    120 &   1.47 &    8$\times$180.0 &   1440.0 &    8,  41,  64, 106 \\
UHW~J247.19598$-$24.47051 & 2008 Jul 2   &  LowRes15  &   0.5$\times$15 &    120 &   1.77 &   10$\times$15.0  &    150.0 &  122, 211, 192, 153 \\
UHW~J247.20293$-$24.44216 & 2008 Jul 2   &  LowRes15  &   0.5$\times$15 &    120 &   1.84 &   11$\times$30.0  &    330.0 &  104, 189, 159, 123 \\
UHW~J247.20293$-$24.55216 & 2008 Jun 27   &  LowRes15  &   0.5$\times$15 &    120 &   1.42 &    10$\times$15.0  &      150.0 &   26,  53,  45,  34 \\
UHW~J247.38539$-$24.04299 & 2011 May 23   &  LowRes15  &   0.5$\times$15 &    120 &   1.86 &    9$\times$180.0 &   1620.0 &   17,  38,  31,  32 \\
UHW~J247.94088$-$24.62855 & 2007 Jun 7   &   ShortXD  &   0.8$\times$15 &    750 &   1.44 &   12$\times$60.0  &    720.0 &   25,  96, 151, 214 \\
UHW~J247.95427$-$24.78060 & 2011 May 23   &  LowRes15  &   0.5$\times$15 &    120 &   1.53 &   12$\times$180.0 &   2160.0 &   10,  29,  23,  30 \\
UHW~J248.69073$-$24.82275 & 2008 Jul 2   &  LowRes15  &   0.5$\times$15 &    120 &   1.56 &   11$\times$180.0 &   1980.0 &   70,  99,  79,  52 \\
\enddata

\tablecomments{The last column gives the median S/N of the spectrum within the standard IR bandpasses.}

\end{deluxetable}

\newpage

\startlongtable
\rotate
\begin{deluxetable}{lllllllll}
\tablecaption{Classification of Candidate Members \label{tbl:spts}}
\tabletypesize{\scriptsize}
\tablewidth{0pt}
\tablehead{
 \colhead{Object} &
 \colhead{SpT} &
 \colhead{$A_V$} &
 \colhead{Gravity} &
 \colhead{Lit SpT} &
 \colhead{Lit Ref}  &
 \colhead{DR2 plx} &
 \colhead{DR2 pmra} &
 \colhead{DR2 pmdec} 
 }
\startdata
\multicolumn{6}{c}{NGC~1333} & 3.38$\pm$0.32 & 7.34$\pm$0.92 & -9.90$\pm$0.6\\
\tableline
UHW~J052.25592+30.82075 &       M7 &  0.00$\pm$0.40 &       {\sc    vl-g} & & &   3.68$\pm$0.23 &   6.72$\pm$0.30 & -11.22$\pm$0.25\\
UHW~J052.26618+30.94156 &  \nodata &  \nodata &  \nodata & &  &  12.62$\pm$1.95 &  11.90$\pm$2.32 &  -3.36$\pm$1.84\\
UHW~J052.27886+31.49911 &       M8 &  4.63$\pm$0.76 &       {\sc    vl-g} & M7, M7-M9 & S12, L16 & & &\\
UHW~J052.30643+31.39637 &  \nodata & \nodata &  \nodata & & & & & \\
UHW~J052.35191+31.47077 &       M6 &  3.33$\pm$0.64 &       {\sc    vl-g} & M7.4, M7, M5 & W04, S12, L16&   3.63$\pm$0.35 &   6.00$\pm$0.57 &  -9.06$\pm$0.37\\
UHW~J052.40147+31.29717 &       L0 &  0.00$\pm$1.69 &       {\sc    vl-g} & late-M, M9& S09, L16& & &\\
UHW~J052.43392+31.32982 &       M8 & 5.66$\pm$2.13\tablenotemark{a} &       {\sc    vl-g} & $\sim$M9, M6-M8.5 & S12, L16& & &\\
UHW~J052.75599+30.63090 &       L0 &  0.00$\pm$1.32 &       {\sc    vl-g} & & & & &\\
\tableline
\multicolumn{6}{c}{IC~348} & 3.09$\pm$0.26 & 4.35$\pm$0.24 & -6.76$\pm$0.52 \\
\tableline
UHW~J055.48646+31.83838 &       M5 &  0.92$\pm$0.46 &  \nodata & & &   7.08$\pm$0.38 & -24.69$\pm$0.53 &  -6.45$\pm$0.33\\
UHW~J055.54657+31.94999 &  \nodata &  \nodata &  \nodata & & &   1.52$\pm$1.74 &   0.83$\pm$2.88 &  -2.55$\pm$1.77\\
UHW~J055.55875+31.95973 &  \nodata &  \nodata &  \nodata & & &   0.29$\pm$0.60 &   1.22$\pm$0.90 &  -1.82$\pm$0.55\\
UHW~J055.67611+31.72142 &       M8 &  5.85$\pm$1.17 &      {\sc    int-g} & & & & &\\
UHW~J055.71237+31.94127 &       M7 &  2.54$\pm$1.17 &      {\sc    int-g} & & & & &\\
UHW~J055.73277+32.07455 &       M7 &  1.23$\pm$1.15 &       {\sc    vl-g} & & &   4.47$\pm$0.66 &   3.96$\pm$1.15 &  -6.30$\pm$0.67\\
UHW~J055.75603+32.03731 &       M6 &  2.73$\pm$0.64 &       {\sc    vl-g} & M5.25 & L16&   3.58$\pm$0.43 &   5.70$\pm$0.87 &  -7.67$\pm$0.57\\
UHW~J055.75810+31.74320 &       M7 &  4.55$\pm$0.54 &       {\sc    vl-g} & & &   2.85$\pm$1.05 &   1.39$\pm$1.32 &  -9.35$\pm$1.00\\
UHW~J055.76964+32.16007 &       M8 &  3.97$\pm$1.17 &       {\sc    vl-g} & & & & &\\
UHW~J055.98976+32.24691 &       M7 &  1.17$\pm$0.54 &       {\sc    vl-g} & M6 & L16&   2.22$\pm$0.61 &   4.07$\pm$1.18 &  -6.41$\pm$0.6\\
UHW~J056.01221+31.87436 &       M6 &  4.68$\pm$0.64 &       {\sc    vl-g} & M6 & L16&   3.19$\pm$0.81 &   4.59$\pm$1.28 &  -6.37$\pm$0.83\\
UHW~J056.04265+31.54385 &       M8 &  0.94$\pm$1.01 &       {\sc    vl-g} & & & & &\\
UHW~J056.07487+31.58792 &  \nodata &  \nodata &  \nodata & & & & &\\
UHW~J056.14403+31.71225 &       M6 &  5.94$\pm$0.51 &       {\sc    vl-g} & & &   5.11$\pm$1.29 &   4.65$\pm$2.31 &  -7.30$\pm$1.48\\
UHW~J056.17321+32.17758 &       M6 &  1.98$\pm$0.51 &       {\sc    vl-g} & M5.75 & L98&   4.37$\pm$0.84 &   4.43$\pm$1.44 &  -6.68$\pm$0.86\\
UHW~J056.34957+32.19849 &       M7 &  2.51$\pm$0.54 &       {\sc    vl-g} & M5.75, M6.5 & A13, L16&   3.92$\pm$0.96 &   2.57$\pm$1.62 &  -7.20$\pm$1.18\\
UHW~J056.37556+31.62740 &  \nodata &  \nodata &  \nodata & & & & &\\
UHW~J056.38941+31.76537 &       M6 &  7.92$\pm$0.51 &       {\sc    vl-g} & & &   3.06$\pm$1.66 &   8.00$\pm$3.41 &  -2.85$\pm$1.81\\
\tableline
\multicolumn{6}{c}{Ophiuchus} & 7.1$\pm$0.4 & -7.2$\pm$2.0 & -25.5$\pm$1.7\\
\tableline
UHW~J245.25929$-$23.97766 &       M6 &  3.86$\pm$0.51 &      {\sc    int-g} & M5.25 & L12&   7.43$\pm$0.57 & -13.11$\pm$1.18 & -25.30$\pm$0.77\\
UHW~J245.32623$-$23.90277 &  \nodata &  \nodata &  \nodata & & & & &\\
UHW~J245.39963$-$23.91766 &       M6 &  2.02$\pm$0.64 &       {\sc    vl-g} & M6 & S06&   7.42$\pm$0.28 & -12.31$\pm$0.57 & -24.36$\pm$0.37\\
UHW~J245.48320$-$23.35050 &  \nodata &  \nodata &  \nodata & & &   7.14$\pm$0.14 & -11.23$\pm$0.29 & -24.28$\pm$0.17\\
UHW~J245.58499$-$23.13122 &       L0 &  2.11$\pm$2.07 &       {\sc    vl-g} & & & & &\\
UHW~J245.66849$-$23.56412 &  \nodata &  \nodata &  \nodata & & &   7.19$\pm$0.22 &  -9.61$\pm$0.48 & -25.75$\pm$0.3\\
UHW~J245.68198$-$24.03424 &       M7 &  5.29$\pm$0.41 &       {\sc    vl-g} & & & & &\\
UHW~J245.71140$-$24.22963 &       M5 &  0.63$\pm$0.31 &  \nodata & & &   7.63$\pm$0.40 & -20.18$\pm$0.77 & -27.58$\pm$0.55\\
UHW~J245.81343$-$23.78480 &       M8 &  3.98$\pm$1.17 &       {\sc    vl-g} & & &   7.86$\pm$1.28 &  -0.77$\pm$3.41 & -23.96$\pm$2.19\\
UHW~J245.87762$-$24.22221 &       M6 &  3.14$\pm$0.64 &       {\sc    vl-g} & & &   6.80$\pm$0.17 &  -7.47$\pm$0.31 & -24.66$\pm$0.19\\
UHW~J245.95168$-$24.22649 &       M7 &  6.70$\pm$1.27 &       {\sc    vl-g} & & &   4.96$\pm$0.50 &  -5.65$\pm$0.89 & -33.75$\pm$0.61\\
UHW~J246.06631$-$24.18753 &       M6 &  5.99$\pm$0.51 &       {\sc    vl-g} & & &   6.57$\pm$0.36 &  -6.13$\pm$0.79 & -23.60$\pm$0.53\\
UHW~J246.51367$-$24.50719 &       L1 &  4.56$\pm$4.40 &       {\sc    vl-g} & L0, M9.7& A12, M12& & &\\
UHW~J246.57742$-$24.49763 &       M9 & 13.21$\pm$3.18 &       {\sc    vl-g} & M6.75, M8.6& A12, M12 & & &\\
UHW~J246.62662$-$24.20326 &       M6 & 19.84$\pm$1.27 &       {\sc    vl-g} & & & & &\\
UHW~J246.65753$-$24.65089 &       M6 &  7.79$\pm$0.51 &       {\sc    vl-g} & M5.25, M6.4& A12, M12&   7.07$\pm$1.07 &  -7.59$\pm$2.40 & -25.48$\pm$1.77\\
UHW~J246.68094$-$24.17893 &       M6 &  4.07$\pm$0.51 &       {\sc    vl-g} & & &   7.66$\pm$0.27 & -10.00$\pm$0.55 & -27.03$\pm$0.35\\
UHW~J246.71774$-$24.02967 &       M7 &  7.32$\pm$1.17 &       {\sc    vl-g} & M7.5 & A10 & & &\\
UHW~J246.73470$-$24.71048 &       L0 &  1.83$\pm$1.87 &       {\sc    vl-g} & M9 & G11& & &\\
UHW~J246.82722$-$24.12273 &       M6 &  7.38$\pm$0.51 &      {\sc    int-g} & M7.5 & A12&   6.76$\pm$1.29 &  -3.80$\pm$3.24 & -27.21$\pm$1.78\\
UHW~J246.90062$-$24.08269 &       M6 &  8.22$\pm$0.51 &       {\sc    vl-g} & & &   6.43$\pm$0.65 &  -9.02$\pm$1.42 & -23.89$\pm$0.84\\
UHW~J246.92017$-$24.48353 &       M9 &  2.49$\pm$1.96 &       {\sc    vl-g} & M7.75, M8.2& A12, M12&   7.14$\pm$0.70 &  -9.46$\pm$1.42 & -25.99$\pm$0.96\\
UHW~J246.94785$-$24.82865 &  \nodata &  \nodata &  \nodata & & & & &\\
UHW~J246.99312$-$23.98747 &       M8 &  2.93$\pm$1.17 &       {\sc    vl-g} & & &   5.46$\pm$0.60 &  -6.83$\pm$1.18 & -25.08$\pm$0.76\\
UHW~J247.01994$-$24.84319 &       M6 &  5.35$\pm$0.76 &      {\sc    int-g} & & &   5.99$\pm$0.41 &  -6.59$\pm$0.88 & -25.22$\pm$0.65\\
UHW~J247.04356$-$24.40568 &       L0 &  7.51$\pm$2.07 &       {\sc    vl-g} & L0 & A12 & & &\\
UHW~J247.19598$-$24.47051 &       M6 &  3.19$\pm$0.64 &       {\sc    vl-g} & M6 & S08&   7.12$\pm$0.20 &  -9.52$\pm$0.39 & -26.44$\pm$0.26\\
UHW~J247.20293$-$24.44216 &       M7 &  1.60$\pm$1.17 &       {\sc    vl-g} & M6.25& A12&   6.26$\pm$0.50 &  -3.93$\pm$0.88 & -26.74$\pm$0.61\\
UHW~J247.20293$-$24.55216 &       M7 &  0.89$\pm$0.96 &       {\sc    vl-g} & & & & &\\
UHW~J247.38539$-$24.04299 &       M6 &  6.42$\pm$0.41 &      {\sc    fld-g} & & & & &\\
UHW~J247.94088$-$24.62855 &       M8 &  5.38$\pm$2.37 &       {\sc    vl-g} & & & & &\\
UHW~J247.95427$-$24.78060 &       L1 &  0.00$\pm$1.25 &       {\sc    vl-g} & & & & &\\
UHW~J248.69073$-$24.82275 &  \nodata &  \nodata &  \nodata & & &  -0.33$\pm$0.38 &   1.02$\pm$0.74 &  -3.50$\pm$0.51\\ 
\enddata
\tablecomments{Mean and standard deviation for Gaia DR2 parallaxes and proper motions are from \citet{ortiz18} for IC~348 and NGC~1333 and \citet{canovas19} for Ophiuchius.}
\tablenotetext{a}{Reddening determined from $J-H$ color rather than $J-K$, as $K$-band flux is anomalously high relative to $J$ and $H$.  \citet{luhman16} interpret this as being scattered light from an edge-on disk.}
\tablerefs{ 
A10: \citet{alves10};
A12: \citet{alves12};
A13: \citet{alves13};
G11: \citet{geers11};
L98: \citet{luhman98};
L12: \citet{luhman12};
L16: \citet{luhman16};
M12: \citet{muzic12};
S06: \citet{slesnick06};
S08: \citet{slesnick08};
S09: \citet{scholz09};
S12: \citet{scholz12};
W04: \citet{wilking04}
}
\end{deluxetable}


\begin{thebibliography}{}

\bibitem[Allers et al.(2006)]{allers06} Allers, K.~N., Kessler-Silacci,
  J.~E., Cieza, L.~A., et al.\ 2006, \apj, 644, 364

\bibitem[Allers \& Liu(2013)]{allers13} Allers, K.~N., \& Liu, M.~C.\
  2013, \apj, 772, 79

\bibitem[Allers et al.(2009)]{allers09} Allers, K.~N., Liu, 
M.~C., Shkolnik, E., et al.\ 2009, \apj, 697, 824 

\bibitem[Alves de Oliveira et al.(2010)]{alves10} 
Alves de Oliveira, C., Moraux, E., Bouvier, J., et al.\ 2010, \aap, 515, A75 

\bibitem[Alves de Oliveira et al.(2012)]{alves12} 
Alves de Oliveira, C., Moraux, E., Bouvier, J., \& Bouy, H.\ 2012, \aap, 539, A151 

\bibitem[Alves de Oliveira et al.(2013)]{alves13} 
Alves de Oliveira, C., Moraux, E., Bouvier, J., et al.\ 2013, \aap, 549, A123   

\bibitem[Barsony et al.(2012)]{barsony12} 
Barsony, M., Haisch, K.~E., Marsh, K.~A., \& McCarthy, C.\ 2012, \apj, 751, 22 

\bibitem[Basri et al.(1996)]{basri96} Basri, G., Marcy, G.~W., \& Graham, J.~R.\ 1996, \apj, 458, 600

\bibitem[Bertin, \& Arnouts(1996)]{sextractor} Bertin, E., \& Arnouts, S.\ 1996, \aaps, 117, 393

\bibitem[Bertin et al.(2002)]{swarp} Bertin, E., Mellier, Y., Radovich, M., et al.\ 2002, adass, 281, 228

\bibitem[Bertin(2006)]{scamp} Bertin, E.\ 2006, adass XV, 351, 112
  
\bibitem[{{Bessell}(2005)}]{2005ARA&A..43..293B}
{Bessell}, M.~S. 2005, ARA\&A, 43, 293

\bibitem[Bohlin(2007)]{bohlin07} Bohlin, R.~C.\ 2007, ASPC 364, The Future 
of Photometric, Spectrophotometric and Polarimetric Standardization, ed. C. Sterken (San Francisco, CA: ASP), 
315 

\bibitem[Brice{\~n}o et al.(2002)]{briceno02} Brice{\~n}o, C., 
Luhman, K.~L., Hartmann, L., Stauffer, J.~R., \& Kirkpatrick, J.~D.\ 2002, 
\apj, 580, 317 

\bibitem[Burgasser(2014)]{burgasser14} Burgasser, A.~J.\ 2014 arXiv:1406.4887

\bibitem[C{\'a}novas et al.(2019)]{canovas19} C{\'a}novas, H., Cantero, C., Cieza, L., et al.\ 2019, \aap, 626, A80

\bibitem[Casali et al.(2007)]{casali07} Casali, M., Adamson, A., Alves de Oliveira, C., et al.\ 2007, \aap, 467, 777

\bibitem[Cushing et al.(2005)]{cushing05} Cushing, M.~C., Rayner, 
J.~T., \& Vacca, W.~D.\ 2005, \apj, 623, 1115 

\bibitem[Cushing et al.(2004)]{cushing04} Cushing, M.~C., Vacca, W.~D., \& Rayner, J.~T.\ 2004, \pasp, 116, 362 

\bibitem[{{Faherty} {et~al.}(2016){Faherty}, {Riedel}, {Cruz}, {Gagne},
  {Filippazzo}, {Lambrides}, {Fica}, {Weinberger}, {Thorstensen}, {Tinney},
  {Baldassare}, {Lemonier}, \& {Rice}}]{2016ApJS..225...10F}
Faherty, J.~K., Riedel, A.~R., Cruz, K.~L., et al.\ 2016, \apjs, 225, 10

\bibitem[Fiorucci \& Munari(2003)]{fiorucci03} Fiorucci, M., \& Munari, U.\ 2003, \aap, 401, 781 

\bibitem[Fitzpatrick(1999)]{fitzpatrick99} Fitzpatrick, E.~L.\ 1999, 
\pasp, 111, 63 

\bibitem[Geers et al.(2011)]{geers11} Geers, V., Scholz, A., Jayawardhana, R., et al.\ 2011, \apj, 726, 23

\bibitem[Girardi et al.(2005)]{trilegal} Girardi, L., Groenewegen, M.~A.~T., Hatziminaoglou, E., et al.\ 2005, \aap, 436, 895

\bibitem[Hambly et al.(2008)]{hambly08} Hambly, N.~C., Collins, R.~S., Cross, N.~J.~G., et al.\ 2008, \mnras, 384, 637

\bibitem[Herczeg, \& Hillenbrand(2014)]{herczeg14} Herczeg, G.~J., \& Hillenbrand, L.~A.\ 2014, \apj, 786, 97

\bibitem[Hewett et al.(2006)]{hewett06} Hewett, P.~C., Warren, S.~J., Leggett, S.~K., et al.\ 2006, \mnras, 367, 454

\bibitem[Irwin et al.(2004)]{irwin04} Irwin, M.~J., Lewis, J., Hodgkin, S., et al.\ 2004, \procspie, 5493, 411
  
\bibitem[{Johnson \& {Morgan}(1953)}]{1953ApJ...117..313J}
Johnson, H.~L., \& {Morgan}, W.~W. 1953, \apj, 117, 313

\bibitem[Jose et al.(2020)]{jose20} Jose, J., Biller, B.~A., Albert, L., et al.\ 2020, \apj, 892, 122

\bibitem[Kirkpatrick et al.(2010)]{kirkpatrick10} Kirkpatrick, J.~D., 
Looper, D.~L., Burgasser, A.~J., et al.\ 2010, \apjs, 190, 100 

\bibitem[Leggett et al.(2006)]{leggett06} Leggett, S.~K., Currie, M.~J., Varricatt, W.~P., et al.\ 2006, \mnras, 373, 781
  
\bibitem[{Liu {et~al.}(2013)Liu, {Dupuy}, \& {Allers}}]{2013AN....334...85L}
Liu, M.~C., {Dupuy}, T.~J., \& {Allers}, K.~N. 2013, AN,
  334, 85

\bibitem[{{Liu} {et~al.}(2016){Liu}, {Dupuy}, \&
  {Allers}}]{2016ApJ...833...96L}
{Liu}, M.~C., {Dupuy}, T.~J., \& {Allers}, K.~N. 2016, \apj, 833, 96

\bibitem[{{Liu} {et~al.}(2003){Liu}, {Najita}, \&
  {Tokunaga}}]{2003ApJ...585..372L}
{Liu}, M.~C., {Najita}, J., \& {Tokunaga}, A.~T. 2003, \apj, 585, 372

\bibitem[{{Lodieu} {et~al.}(2013){Lodieu}, {Dobbie}, {Cross}, {Hambly}, {Read},
  {Blake}, \& {Floyd}}]{2013MNRAS.435.2474L}
{Lodieu}, N., {Dobbie}, P.~D., {Cross}, N.~J.~G., {Hambly}, N.~C., {Read},
M.~A. 2013, \mnras, 435, 2474

\bibitem[Lodieu et al.(2018)]{2018MNRAS.473.2020L} Lodieu, N., Zapatero Osorio, M.~R., B{\'e}jar, V.~J.~S., et al.\ 2018, \mnras, 473, 2020

\bibitem[Lord(1992)]{lord92}
Lord, S.D. 1992, NASA Technical Memor. 103957

\bibitem[Luhman et al.(2016)]{luhman16} Luhman, K.~L., Esplin, T.~L., \& Loutrel, N.~P.\ 2016, \apj, 827, 52 

\bibitem[Luhman, \& Mamajek(2012)]{luhman12} Luhman, K.~L., \& Mamajek, E.~E.\ 2012, \apj, 758, 31

\bibitem[Luhman et al.(1998)]{luhman98} Luhman, K.~L., Rieke, G.~H., Lada, C.~J., et al.\ 1998, \apj, 508, 347

\bibitem[{Mainzer \& {McLean}(2003)}]{2003ApJ...597..555M}
Mainzer, A.~K., \& {McLean}, I.~S. 2003, \apj, 597, 555

\bibitem[Mart{\'\i}n et al.(1996)]{1996ApJ...469..706M} Mart{\'\i}n,
  E.~L., Rebolo, R. \& Zapatero-Osorio, M. 1996, \apj, 469, 706

\bibitem[Mart{\'\i}n et al.(1998)]{martin98} Mart{\'\i}n, E.~L., Basri,
  G., Gallegos, J.~E., et al.\ 1998, \apjl, 499, L61
  
\bibitem[Megessier(1995)]{megessier95} Megessier, C.\ 1995, \aap, 296, 771 

\bibitem[Muench et al.(2007)]{muench07} Muench, A.~A., Lada, 
C.~J., Luhman, K.~L., Muzerolle, J., \& Young, E.\ 2007, \aj, 134, 411 

\bibitem[Mu{\v z}i{\'c} et al.(2012)]{muzic12} Mu{\v z}i{\'c}, 
K., Scholz, A., Geers, V., Jayawardhana, R., 
\& Tamura, M.\ 2012, \apj, 744, 134 

\bibitem[Najita et al.(2000)]{najita00} Najita, J.~R., Tiede, 
G.~P., \& Carr, J.~S.\ 2000, \apj, 541, 977

\bibitem[Ortiz-Le{\'o}n et al.(2018)]{ortiz18} Ortiz-Le{\'o}n, G.~N., Loinard, L., Dzib, S.~A., et al.\ 2018, \apj, 865, 73

\bibitem[Pe{\~n}a Ram{\'\i}rez et al.(2012)]{2012ApJ...754...30P} Pe{\~n}a Ram{\'\i}rez, K., B{\'e}jar, V.~J.~S., Zapatero Osorio, M.~R., et al.\ 2012, \apj, 754, 30
  
\bibitem[Rayner et al.(2009)]{rayner09} Rayner, J.~T., Cushing, M.~C., \& Vacca, W.~D.\ 2009, \apjs, 185, 289 

\bibitem[Rayner et al.(2003)]{rayner03} Rayner, J.~T., Toomey, D.~W., Onaka, P.~M., et al.\ 2003, \pasp, 115, 362 

\bibitem[Rebolo et al.(1998)]{1998Sci...282.1309R} Rebolo, R., Zapatero Osorio, M.~R., Madruga, S., et al.\ 1998, Sci, 282, 1309
  
\bibitem[Robin et al.(2003)]{robin03} Robin, A.~C., Reyl{\'e}, C., Derri{\`e}re, S., \& Picaud, S.\ 2003, \aap, 409, 523 
  
\bibitem[{{Rosenthal} {et~al.}(1996){Rosenthal}, {Gurwell}, \&
  {Ho}}]{1996Natur.384..243R}
{Rosenthal}, E.~D., {Gurwell}, M.~A., \& {Ho}, P. T.~P. 1996, \nat, 384, 243

\bibitem[Scholz et al.(2009)]{scholz09} Scholz, A., Geers, V., Jayawardhana, R., et al.\ 2009, \apj, 702, 805

\bibitem[Scholz et al.(2012)]{scholz12} Scholz, A., 
Jayawardhana, R., Muzic, K., et al.\ 2012, \apj, 756, 24 

\bibitem[Simons \& Tokunaga(2002)]{simons02} Simons, D.~A., \& Tokunaga, A.\ 2002, \pasp, 114, 169

\bibitem[Skrutskie et al.(2006)]{skrutskie06} Skrutskie, M.~F., Cutri, R.~M., Stiening, R., et al.\ 2006, \aj, 131, 1163

\bibitem[Slesnick et al.(2006)]{slesnick06} Slesnick, C.~L., Carpenter, J.~M., \& Hillenbrand, L.~A.\ 2006, \aj, 131, 3016

\bibitem[Slesnick et al.(2008)]{slesnick08} Slesnick, C.~L., Hillenbrand, L.~A., \& Carpenter, J.~M.\ 2008, \apj, 688, 377
  
\bibitem[{{Str{\"o}mgren}(1966)}]{1966ARA&A...4..433S}
{Str{\"o}mgren}, B. 1966, ARA\&A, 4, 433

\bibitem[Testi(2009)]{testi09} Testi, L.\ 2009, \aap, 503, 639 

\bibitem[{Tinney {et~al.}(2005)Tinney, {Burgasser}, {Kirkpatrick}, \&
  {McElwain}}]{2005AJ....130.2326T}
Tinney, C.~G., {Burgasser}, A.~J., {Kirkpatrick}, J.~D., \& {McElwain}, M.~W.
2005, \aj, 130, 2326

\bibitem[Tokunaga et al.(2002)]{tokunaga02} Tokunaga, A.~T., 
Simons, D.~A., \& Vacca, W.~D.\ 2002, \pasp, 114, 180 

\bibitem[Tokunaga \& Vacca(2005)]{tokunaga05} Tokunaga, A.~T., \& Vacca, W.~D.\ 2005, \pasp, 117, 421 

\bibitem[Vacca et al.(2003)]{vacca03} Vacca, W.~D., Cushing, M.~C., \& Rayner, J.~T.\ 2003, \pasp, 115, 389 

\bibitem[Wilking et al.(2008)]{wilking08} Wilking, B.~A., Gagn{\'e}, M., \& Allen, L.~E.\ 2008, in Handbook of Star Forming Regions, Volume II: The Southern Sky, ed. B. Reipurth (San Francisco, CA: ASP), 351

\bibitem[Wilking et al.(2004)]{wilking04} Wilking, B.~A., Meyer, M.~R., Greene, T.~P., et al.\ 2004, \aj, 127, 1131

\bibitem[Wilking et al.(2005)]{wilking05} Wilking, B.~A., Meyer, 
M.~R., Robinson, J.~G., \& Greene, T.~P.\ 2005, \aj, 130, 1733

\bibitem[Young et al.(2015)]{young15} Young, K.~E., Young, C.~H., Lai, S.-P., et al.\ 2015, \aj, 150, 40

\bibitem[{{Zapatero Osorio} {et~al.}(2017){Zapatero Osorio}, {B{\'e}jar}, \&
  {Pe{\~n}a Ram{\'\i}rez}}]{2017ApJ...842...65Z}
{Zapatero Osorio}, M.~R., {B{\'e}jar}, V.~J.~S., \& {Pe{\~n}a Ram{\'\i}rez}, K.
2017, \apj, 842, 65

\end{thebibliography}
\end{document}